\documentstyle[12pt]{article}
\begin{document}
\sf

\begin{center}
{\LARGE \bf PION CLOUD CONTRIBUTION TO \\
\vspace{0.3cm}
$K^{+}$ NUCLEUS SCATTERING}
\end{center}

\begin{center}
{\large C. Garc\'{\i}a-Recio$^{(1)}$, J. Nieves$^{(2)}$ and
E. Oset$^{(3)}$}
\end{center}

\noindent
{\small $^{(1)}$ Dpto. de F\'{\i}sica Moderna,
Facultad de Ciencias, Universidad de Granada,\\ E-18071
Granada, Spain.}

\noindent
{\small $^{(2)}$ Physics Department,
The University, Southampton SO9 5NH, United Kingdom.}

\noindent
{\small $^{(3)}$ Dpto. de F\'{\i}sica Te\'orica and
IFIC, Centro Mixto Universidad de Valencia-CSIC.
E-46100 Burjassot, (Valencia), Spain.}

\vspace{1cm}

\begin{abstract}
A careful reanalysis is done of the contribution to $K^{+}$ nucleus
scattering from the interaction of the kaon with the virtual pion
cloud. The usual approximations made in the evaluation of the related
kaon selfenergy are shown to fail badly. We also find new interaction
mechanisms which provide appreciable corrections to the kaon
selfenergy. Some of these contribute to the imaginary part below pion
creation threshold. The inclusion of these new mechanisms in the
inelastic part of the optical potential produces a significant
improvement in the differential and total $K^{+}$ nuclear cross
sections. Uncertainties remain in the dispersive part of the optical
potential.
\end{abstract}

\vspace{2cm}
Submitted to Physical Review C

\vspace{2cm}
Preprint number: UG-DFM-2/94
\newpage

\noindent
{\large \bf 1.- Introduction}

Systematic discrepancies between the microscopic optical potential
calculations  for
$K^{+}$  nucleus scattering \cite{1,2,3} and the experimental data
\cite{4,5,6,7,8} have led to suggestions that it may be an indication
of an increased size of the nucleons in the nucleus \cite{2,9,10}.
These discrepancies remain when a number of conventional nuclear
corrections (Pauli blocking, nucleon-nucleon correlations, off-shell
corrections, etc) are taken into account (\cite{2}-\cite{3}).
Parallelly,
work has been done about the
contribution of the nuclear pion cloud to the $K^{+}$ optical
potential \cite{11,12}. In \cite{11} a qualitative estimate is done
of the meson cloud effects by assuming that the $K^{+}N$ cross
section is increased by $\delta n_{\pi} \sigma (K^{+} \pi)$, where
$\delta n_{\pi}$ is the excess number of pions per nucleon
in the nucleus.
In
ref. \cite{12}, together with a good summary of the status of the
problem,  a thorough and instructive study of the meson cloud
contribution to the scattering amplitude is done by evaluating
explicitly the real and the imaginary parts. A $K^{+} \pi$ amplitude
with off-shell extrapolation and crossing symmetry, inspired in the
work on the $\pi \pi$ interaction \cite{13} is used. The work relies
upon the pion excess distribution found in \cite{14}, which accounts
for $ph$ and $\Delta h$ components in a correlated ground state. The
interference of $ph$ and $\Delta h$ components is essential to
produce a positive pion excess number in the nucleus \cite{14,ref15}.

In the present work we have made a more rigorous evaluation of the
pion cloud contribution to the $K^+$ nucleus optical potential which requires
only the knowledge of the pion propagator in the nuclear medium and
a realistic model for the $K\pi$ amplitude.

In the nuclear medium, the pion propagator is renormalized by allowing the
pion excite $ph$ and $\Delta h$ components, such a model provides a
realistic model for the $\pi$ nucleus interaction and accounts for
the basic components needed to produce realistic pion numbers in
finite nuclei \cite{ref15}. Our model for the pion propagator is briefly
described in Appendix A. The pion distribution is not needed
explicitly in our computation of the
$K^+$-nucleus optical potential, although the formal connection of the pion
propagator
to the  pion distribution, $n ({\bf q})$, will be made. Actually, one
of our findings is that the pion cloud contribution to the imaginary
part of the $K^{+}$ - nucleus optical potential can not be cast as an integral
of the form $\int d^{3}{\bf q} ~ n({\bf q})~ f({\bf q})$ as assumed
in \cite{12} and also implicitly in \cite{11}.

For the $K\pi$ amplitude we use the model of ref. \cite{12}. This
model incorporates on-shell conditions and crossing symmetry. A
detailed study is made in ref. \cite{12} about uncertainties from the
off-shell extrapolation, form factors, etc., allowing us to simplify
the discussions and concentrate on the novelties that the present
work introduces. For the sake of completeness, in Appendix B this
$K\pi$ model is summarized.

On the other hand, we introduce new mechanisms also related to
the scattering of positive kaons with the pion cloud, which have not been
considered previously and are found to be very important.

The calculations are done in infinite nuclear matter and the
contributions to the $K^{+}$ selfenergy are obtained as a function of
$\rho$, the nuclear matter density. By means of the local density
approximation, carefully studied and justified in \cite{ref16} in
connection with $\pi$-nucleus scattering, we obtain the meson
exchange currents (MEC) contribution to the $K$-nucleus optical
potential as a function of $\rho (r)$. Our model for the
$K$-nucleus potential is obtained by adding these new
contributions, calculated in the present work,
 to the
conventional ones from the impulse approximation (see Appendix C)
and to the standard nuclear corrections (nuclear correlations,
off-shell and binding effect, Pauli exclusion,... calculated
in refs.~\cite{2,6}). This new optical potential is then used
to obtain the differential and total $K^+$-nucleus cross sections by
solving numerically the Klein Gordon equation. In the
following, we will refer
indistinctly to the kaon selfenergy or to the optical potential, as
they are related by $\Pi(k)=2k^0V_{\rm opt}(k)$. Nevertheless it is
the selfenergy what appears in the Klein-Gordon equation.

The paper is organized as follows: In order to test the model used for pions
in nuclei,
the excess number of pions in the nucleus is calculated in section 2.
In section 3  we recalculate the contribution from the
MEC mechanism considered in refs.~\cite{11,12} but relaxing the
static approximation. The new MEC
mechanisms contributing to the the $K^+$ selfenergy in nuclei are
presented in section 4 where is also evaluated their contribution
to the imaginary part of the $K^+$-nucleus optical potential.
Section 5 is devoted to the study of the MEC
contribution to the real part of $K^+$ selfenergy. In
section 6, the $K^+$-nucleus differential and total cross-sections
calculated with IA, with IA+MEC, and with the conventional optical
potential plus MEC, are shown and compared with
experimental data for $^{12}$C and $^{40}$Ca and also with the ratio
of those cross-sections over the $K^+$ deuterium cross-section.
Finally, in section 7, we summarize and parametrize the results for the MEC
contribution to the $K^+$ optical potential and present our conclusions.

\vspace{1cm}

\noindent
{\large \bf 2.- The pion propagator and the number of pions.}

\nobreak
The propagator of a pion with four-momentum $q$ (and isospin
$\lambda$) should satisfy the Lehmann representation \cite{ref17}
\begin{eqnarray}
D (q) = \int_{0}^{\infty} {d \omega\over \pi} ~(- 2 \omega)
\frac{{\rm Im}D(\omega, {\bf q})}{ {q^0}^2 - \omega^{2}
+ i \epsilon} \ \ ,
\label{e1}
\end{eqnarray}
\noindent
Our calculation of the pion propagator in Appendix A
preserves the appropriate analytical properties and
therefore eq.~(\ref{e1}) holds.

Another equation which is relevant
in connection with the present problem is the sum rule
\begin{eqnarray}
- \int_{0}^{\infty} \frac{dq^{0}}{\pi} ~2 q^{0}~ {\rm Im}D (q^{0},
{\bf q}) = 1.
\label{e2}
\end{eqnarray}
This equation expresses the equal time commutation relation of the
pion fields. We check that eq.~(\ref{e2}) is fulfilled in our model
at the level of one per thousand which is sufficient for our
purposes.

A check of the model for pion propagation in the nucleus is to
calculate the number of pions it produces. Although this quantity  is
not needed in our evaluation of the pion cloud contribution to the
$K^+$ nucleus scattering, we show our results for it in order to
compare with earlier work. Let $n_\lambda({\bf q})$ be the pion
number distribution for a single class of pions and $n({\bf q})$ the
total number of pions, namely
\begin{eqnarray}
n ({\bf q} \,) \equiv\sum_\lambda n_\lambda ({\bf q} \,) =
\sum_\lambda\langle
 a^{+}_{{\bf q}\lambda} a_{{\bf q}\lambda}
\rangle,
\label{e3}
\end{eqnarray}
where the symbol $\langle~\rangle$ indicates the expectation value
in the nuclear ground state and $a_{{\bf q}\lambda}$ the
annihilation operator of a pion with momentum ${\bf q}$ and isospin
$\lambda$.  Thus,
\begin{eqnarray}
\int \frac{d^{3} {\bf q}}{(2 \pi)^{3}}~  n ({\bf q}) =
\frac{n_{\pi}}{V} = \frac{n_{\pi}}{A} \rho ,
\label{e4}
\end{eqnarray}
\noindent
with $n_{\pi}/A$ the total number of pions per nucleon. More
amenable to calculation are the quantities $N ({\bf q})$ and
$\delta {N} ({\bf q})$,
\begin{eqnarray}
{N} ({\bf q}) & = &  n ({\bf q}) + \frac{1}{2} \sum_\lambda\langle
 a^{+}_{{\bf q},\lambda}
a_{-{\bf q},-\lambda}^{+}
\rangle + \frac{1}{2} \sum_\lambda\langle
a_{{\bf q},\lambda} a_{-{\bf q},-\lambda}
\rangle,
\label{e5}\\
\delta {N} ({\bf q}) & = & {N} ({\bf q}) -
\rho \left ({\partial {N} ({\bf q})\over
\partial\rho}\right)_{\rho=0}
\label{e6}
\end{eqnarray}
 The linear term in $\rho$ accounts for
the number of
pions per free nucleon, then it is subtracted in eq.~(\ref{e6})
to obtain  the ``pion excess'' $\delta N({\bf q})$.

According to eq.~(\ref{e4}) the following integral
\begin{eqnarray}
\frac{1}{\rho} \int \frac{d^{3} {\bf q}}{(2 \pi)^{3}}
\delta {N} ({\bf q}) \equiv \frac{\delta N_{\pi}}{A} .
\label{e7}
\end{eqnarray}
provides the ``excess number of pions'' per nucleon, counting
the three isospin states and also
$\langle
 a^{+}_{{\bf q},\lambda} a_{-{\bf q},-\lambda}^{+}
\rangle$
and $\langle
 a_{{\bf q},\lambda} a_{-{\bf q},-\lambda}
\rangle$ as it co\-mes from eq.~(\ref{e5}).
Accord\-ing to the results of ref~\cite{ref18} on the contribution of
the pion cloud to Compton scattering, $n_\lambda({\bf q})$ is equal to
${1\over 2} \langle  a^{+}_{{\bf q},\lambda} a_{-{\bf
q},-\lambda}^{+} \rangle +{1\over 2} \langle a_{{\bf q},\lambda}
a_{-{\bf q},-\lambda}\rangle$, then the results of previous papers on
the pion number excess $\delta n({\bf q})$ and $\delta n_\pi$ are to
be compared with $\delta N({\bf q})/2$ and $\delta N_\pi/2$.

These quantities can be computed using the relationships
\begin{eqnarray}
\frac{1}{3}N({\bf q}) &=& -{2 \omega ({\bf q} )}
 \int_{0}^{\infty} \frac{d q^{0}}{2 \pi}
 {\rm Im}[D (q) -D_{0} (q) ]
 \label{Ee8} \\
\frac{1}{3}\delta N({\bf q}) &=& -{2 \omega ({\bf q} )}
 \int_{0}^{\infty} \frac{d q^{0}}{2 \pi}
 {\rm Im}[\delta D (q) ]
 \label{Ee9}
\end{eqnarray}
where $\omega({\bf q})=\sqrt{m_\pi^2+{\bf q}^2}$ and
\begin{eqnarray}
\delta D(q) = D(q)-D_0(q)-\rho\left({\partial D(q)\over\partial\rho}
\right)_{\rho=0}\ \ .
\label{Ee10}
\end{eqnarray}
Eq.~(\ref{Ee8}) can be derived by writing the pion propagator in
terms of the creation and annihilation operators and making use of
eq.~(\ref{e1})

In fig.~{1}, we show $\delta N({\bf q})$ (solid line) for normal
($\rho=\rho_0=0.17$\ fm$^{-3}$) nuclear matter. We have also
calculated  $\delta N({\bf q})$ for different values of the density,
observing that it behaves quadratically in density. If only $ph$
excitations are considered one obtains the dotted line in fig.~{1}.
By integrating it a negative pion excess number is obtained as it was
already pointed out in \cite{12}. If only $\Delta h$ excitations are
considered one obtains the result of the dashed line, which
represents a very small pion excess number. When considering
simultaneously both, $ph$ and $\Delta h$, excitations the solid line
of fig.~{1} is obtained. $\delta N({\bf q})$ is larger for large
${\bf q}$ than before and a positive excess number of pions is found
when integrating over $d^3{\bf q}$, thus proving that the
interference of $ph$ and $\Delta h$ is essential to produce a
positive excess number.

The distribution $\delta {N} ({\bf q})$ has identical shape to the
one  from \cite{14}. For $\rho = \rho_{0}$ the integral of
eq.~(\ref{e7}) gives 0.67, half of it coming from the integral of the
strict pion number, $\delta n({\bf q})$, according to \cite{ref18}.
This gives us 0.33 pions per nucleon in nuclear matter at $\rho =
\rho_{0}$. Since $\int d^{3} {\bf q}~ \delta N ({\bf q})$ is
proportional to $\rho^{2}$ and in a finite nucleus like $^{12}$C the
magnitude ${1\over A}\int d^{3} {\bf r}~ \rho^{2} ({\bf r})$ is
around a factor two smaller than in nuclear matter, then we obtain
0.17 pions per nucleon in the $^{12}$C nucleus, on the upper edge of
the band of values obtained by other authors \cite{14,ref15}.
\vspace{1cm}

\noindent
{\large \bf 3.- Formal derivation of the ``standard'' pion cloud
 contribution}

\nobreak
We call here ``standard'' mechanism, to the one depicted in
fig.~{2}. This is the only one considered in previous papers, and
then it was calculated in the static approach which we describe in
subsection 3.1. In subsection 3.2 we calculate it exactly.

The $K^{+}$ selfenergy of the basic diagram shown in fig.{2}a  in an
infinite spin-isospin symmetric nuclear medium  is given by
\begin{eqnarray}
- i \Pi (k) = \int \frac{d^{4} q}{(2 \pi)^{4}} ~ i D (q)~ (-i)~
\frac{1}{2}~ 3~ t^{0} (k,q;k,q),  \label{Ee11}
\end{eqnarray}
\noindent
where  $t^{0}$ is the isoscalar $K^+ \pi$ amplitude (average of
$t_{K^{+} \pi^{i}}$ for the three charged pions) and the factor
$\frac{1}{2}$ is a symmetry factor. However, as depicted in
figs.~{2b}, {2c}, {2d} the full propagator contains the free pion,
one $ph$ or $\Delta h$ corrections and higher order corrections with
$2ph$, $ph$ $\Delta h$, $2 \Delta h$, etc, excitations. The
contribution from the free pion has to be subtracted because it
corresponds to a piece in the free $K^{+}$ selfenergy. Analogously,
once this subtraction is made, we will have terms in the selfenergy
coming from $1ph$ or 1$\Delta h$ excitations which are proportional
to $\rho$.  These terms must also be subtracted because they are
implicitly accounted for in the IA selfenergy $\Pi^{\rm IA} = t_{KN}~
\rho$, where $t_{KN}$ is the empirical $KN~t$-matrix.  Hence the
genuine pion cloud contribution to the $K^{+}$ selfenergy is given by
\begin{eqnarray}
\delta\Pi (k) = i \int \frac{d^{4} q}{(2 \pi)^{4}}~ \delta D (q)
{}~\frac{3}{2}~ t^{0} (k, q; k, q), \label{Ee12}
\end{eqnarray}
\vspace{0.7cm}
\noindent
{ \bf 3.1.- Static Approximation}

\nobreak
In the static approximation the $q^0$ dependence in the $t$ matrix is
neglected. For instance in \cite{12} $q^0$ is set to zero in $t^{0}
(k,q;k,q)$. In this case one obtains
\begin{eqnarray}
\delta\Pi_{\rm stat} (k) &=& - \int \frac{d^{3} {\bf q}}
{(2 \pi)^{3}} \int^{\infty}_{0}
\frac{d q^{0}}{2 \pi}
{\rm Im}\delta D (q) ~ 3 \left [ t^{0} (k,q;k,q) \right ]_{q^{0}
{\rm fix}}
\nonumber \\
&=&\int {d^3{\bf q}\over
(2\pi)^3} {\delta N({\bf q})\over 2\omega({\bf q})}
{}~ t^{0} (k,q;k,q) \big |_{q^{0}
{\rm fix}} ,
 \label{e13}
\end{eqnarray}
\noindent
where the first equality follows from $\int dq^0~{\rm Re} D(q)~=~0$,
and $D(q^0,{\bf q})=D(-q^0,{\bf q})$.

Hence, in the static approximation $\Pi (k)$ comes as a
weighted integral of the
$K^+\pi $ amplitude with the pion distribution in the nucleus.
This result looks intuitive but recall that $N({\bf q})$
 contains $n({\bf q})$ and also the
expectation values $\langle
 a_{{\bf q}\lambda} \; a_{-{\bf q}-\lambda}
\rangle$
and $\langle
 a^{+}_{{\bf q}\lambda} \; a_{-{\bf q}-\lambda}^{+}
\rangle$.
These three factors correspond in fact to having the $K^+$ scattering
with a pion, annihilating two pions from the ground state or creating
two pions from the ground state, as symbolically depicted in
fig.~{3}. Note that with our  field theoretical formalism, the three
terms are automatically included.

The approach of \cite{12} corresponds to the static approximation of
eq.~(\ref{e13}) with $\delta N({\bf q}) = 2\times \delta n({\bf q})$
and $\delta n({\bf q})$ taken from \cite{14}. The factor 2 accounts
for the two pion creation or annihilation mechanisms as found in
\cite{ref18}. As already noted in \cite{12} these extra terms are
ignored in the approach of \cite{11}.

Our claim here is that the static approximation which justifies the
approaches of \cite{11,12} is inaccurate, particularly for the
imaginary part of $\Pi (k)$.

Indeed, the imaginary part of $\delta\Pi_{\rm stat}$ from
eq.~(\ref{e13}) is given by
\begin{eqnarray}
{\rm Im} \delta\Pi_{\rm stat} (k) = - \int
\frac{d^{3} {\bf q}}{(2 \pi)^{3}} \int_{0}^{\infty}
\frac{d q^{0}}{2 \pi} \,
{\rm Im}\delta D (q) ~3
\left [ {\rm Im} t^{0} (k,q;k,q)\right ]_{q^{0}
{\rm fix}} .
 \label{e14}
\end{eqnarray}
Note that the range of the ${q^0}$ integration goes from 0 to
$\infty$. This will be very different in the exact case which we
analyze below, implying, as we shall see, that the static
approximation is not good.

\vspace{0.7cm}
\noindent
{ \bf 3.2.- Exact Calculation of ${\bf\rm Im}{\bf\delta\Pi}$ for the
``standard'' mechanism}

\nobreak
Now let us find the exact expression for ${\rm Im}\delta\Pi (k)$.
This requires a
knowledge of the analytical structure of  $t^{0} (k,q;k,q)$.
We assume that this amplitude can be written in the following way:
\begin{eqnarray}
t^{0} (k,q;k,q) = \tilde{t} (s) + \tilde{t} (u) ,
 \label{e15}
\end{eqnarray}
\noindent
where $s = (k + q)^{2}, \, u = (k - q)^{2}$ are the usual Mandelstam
variables ($t$= 0 in our case), which automatically satisfies
crossing symmetry. This is the case for the model which we will use.
The function $\tilde {t} (x)$ has the right analytical properties and
develops and imaginary part for $x > x_{0} = (m_{K} + {m_\pi})^{2}$.
It also satisfies the subtracted dispersion relation
\cite{ref17,ref19}
\begin{eqnarray}
\tilde{t} (s) &=& P (s) +(s-x_0)
\int_{x_{0}}^{\infty}  \frac{dx}{\pi}
\frac{{\rm Im}~\tilde{t} (x)}{(s - x + i \epsilon)
(x_{0} - x)} \ ,  \label{e16}
\end{eqnarray}
\noindent
where $P(x)$ is a real polynomial.

Using this form of $t^0$ from eq.~(\ref{e15}) in $\delta\Pi$ of
eq.~(\ref{Ee12}) and the symmetry of the integrand under $
q\leftrightarrow - q$. we obtain:
\begin{eqnarray}
\delta\Pi (k)  =   i 3 \int {d^4 q\over (2\pi)^4}
\delta D(q)\tilde t(u) \label{Ee17}
\end{eqnarray}
The analytical structure of $\delta D (q) \tilde{t} (u)$ in the
variable $q^{0}$ is shown in fig.~{4}. It has cuts and poles in the
second and fourth quadrants from $\delta D(q)$ and two single poles
in $q^{0} =  k^{0} \pm E (x) \mp i \epsilon$, with $E (x) = [
({\bf k} - {\bf q})^{2} + x]^{1/2}$. This particular structure
suggests a Wick rotation, as indicated in fig.~{4}, in order to
perform the $q^{0}$ integral. Since the integral vanishes at the
circles of infinite radius, we have
\begin{eqnarray}
i \int_{-  \infty}^{\infty} dq^0~= i \int_{- i \infty}^{i \infty}
 d q^{0} - 2 \pi~ {\rm Res} ( q^0=k^{0} - E (x)) ~\theta
(k^{0}-E(x)) ,
 \label{Ee18}
\end{eqnarray}
\noindent
The integral over the imaginary axis is real and the only source of
imaginary part comes from the residue at the pole. Thus,
\begin{eqnarray}
\lefteqn{{\rm Im}~ \delta\Pi (k)=}\nonumber \\
& = & - \int \frac{d^{3} {\bf q}}{(2 \pi)^{3}}
\int_{x_{0}}^{\infty} \frac{dx}{\pi}
\theta(k^0-E(x))~
\frac{{\rm Im} \tilde{t} (x)}{2 E (x)} 3
{\rm Im}~\delta D(q) |_{q^{0}
= k^{0} - E (x)} \nonumber \\
& = &
  - \int \frac{d^{3} {\bf q}}{(2 \pi)^{3}}
\theta (k^{0} - E (x_{0}))
\int_{0}^{k^{0} - E (x_{0})}
\frac{d q^{0}}{2\pi}
6{\rm Im}\tilde{t} (u)
{\rm Im}\delta D (q) \label{Ee19} .
\end{eqnarray}
Note that $\tilde{t}$ appears at the end with argument $u$ rather
than $s$. By comparing eq.~(\ref{Ee19}) with the static expression of
eq.~(\ref{e14}) we find a main substantial difference in the fact
that the $q^{0}$ integral goes from 0 to $\infty$ in the static
formula while here it is restricted to the interval $[ 0, k^{0} - E
(x_{0})]$. Hence, even if we make ${\rm Im} \tilde{t} (u)$  static in
eq.~(\ref{Ee19}) in order to take it out of the $q^{0}$ integral, the
pion excess number $\delta N({\bf q})$ will not be generated because
the range $[0, \infty]$ in the $q^{0}$ integration is needed in
eq.~(\ref{Ee9}). Note that the range of ${\bf q}$ is also restricted
because $E(x_0)< k^0$. Then the whole phase space allowed is finite,
as corresponds to the reaction channels accounted for by ${\rm
Im}\Pi(k)$. Under these circumstances one should not expect the
static approximation to provide realistic results.

The pathologies generated by the intuitive use of the particle number
are general in decay processes or in the evaluation of imaginary
parts of amplitudes, i.e., in cases where conservation of energy and
momentum is at stake. This occurs because the relevant magnitude is
${\rm Im}D (q)$ which provides the probability of finding a pion with
momentum ${\bf q}$ and energy $q^{0}$. The probability of finding a
pion of momentum ${\bf q}$ is an integral property obtained when one
integrates over the energy of the pion from 0 to $\infty$. However,
in decay processes the range of energies allowed is limited because
of energy and momentum conservation, and the particle number can not
be factored out. A spectacular example of the failure of the static
approximation has been shown in \cite{ref20} in connection with the
mesonic $\Lambda$ decay in nuclei. The argument goes as follows: The
$\Lambda \rightarrow \pi N$ decay is forbidden in nuclei because the
nucleon momentum, $k_{N} \simeq 100 {\rm MeV} /c$ is below the Fermi
momentum $k_{F} \simeq 270 {\rm MeV} /c$ (because of surface effects
the decay it still possible but appreciable reduced, about 4 orders
of magnitude in heavy nuclei). However, since the occupation number
for states below the Fermi momentum is not 1 but about 0.85, it was
implicitly  argued in \cite{ref21} that the $\Lambda$ mesonic decay
in nuclei should saturate at values about 15$\%$ of the free
$\Lambda$ width [up to a moderate effect of pion absorption in the
nucleus]. The argument, however intuitive, suffers from the same
defects of the static approximation discussed here and it was found
in \cite{ref20} that, the actual results for the mesonic width are
about three orders of magnitude smaller than the results of the
intuitive argument based on the nucleon distribution in nuclei.

The interesting expression for ${\rm Im}\delta\Pi (k)$ of
eq.~(\ref{Ee19}) indicates that one needs only ${\rm Im} D (q)$ and
${\rm Im} \tilde{t} (u)$ to obtain ${\rm Im}\delta\Pi (k)$ and only
in a reduced range of $q^{0}$ and ${\bf q}$. Although one can in
principle evaluate ${\rm Im}\delta\Pi (k)$ from eq.~(\ref{Ee12}), it
is a highly inefficient and dangerous method because of the strong
cancellations and the large ranges of $q^{0}$ involved in the
integrations.

For the real part of $\delta\Pi$ we do not find finite ranges of
integration.

\vspace{0.7cm}
\noindent
{ \bf 3.3.- Results of calculations for Im${\bf\delta\Pi}$}

\nobreak
For the explicit calculations we use the $K^+\pi$ amplitude from
ref.~\cite{12}, which is summarized in Appendix B. Calculations with
different values of $\rho$ have shown that Im$\delta\Pi$ behaves
quadratically in density. In fig.~{5} we present the imaginary part
of the $K^+$ selfenergy calculated for nuclear matter at normal
density $\rho_0$. The dot-dashed line is the result of the static
approximation using $q^0$ fixed to zero. The dashed line displays the
exact result calculated as explained in subsection 3.2. The leading
part of the optical potential comes from the IA. For comparison, it
is displayed in the figure with a crossed solid line. We observe that
the static result is about twice the exact one, and also that in any
case both of them are very small compared to the IA. Then this mechanism is
not enough to account for the experimental results. For comparison we
also show there the results of the total MEC contribution when
considering the new mechanisms which we discuss below.

\vspace{1cm}

\noindent
{\large \bf 4.- New mechanisms from the pion cloud.}

\nobreak

The imaginary part of $\delta\Pi$ of fig.~{2} is related to
Im$\delta D$ and Im$t_{K\pi}$ by eq.~(\ref{Ee19}). This means that
the reactive channels of the $K^+$ selfenergy
are due simultaneously to the
reaction channels of the pion in nuclear matter and the reaction
channels of $t_{K\pi}$.
For the kaon kinetic energies which will be considered in this work
the only open channels in $t_{K\pi}$ are the elastic one or the charge
exchange, $K^+\pi_i\to K^0\pi_j$, and thus Im$t_{K\pi}$ is due only
to the process $K\pi\to K\pi$. Hence, Im$t_{K\pi}$ is related through
the optical theorem to $|t_{K\pi}|^2 {\rm Im}D_0 {\rm Im}D_K$ as
it is given diagrammatically in fig.~{6}. Then, using the optical theorem
in eq.~(\ref{Ee19}), we obtain the identity shown diagrammatically in
fig.~{7}. This allows us to understand the processes to which
Im$\delta\Pi$ is due, these are: $K~N\to~ K~\pi~N,\ \ K~N\to~ K~\pi~
\Delta  $, where the interaction $KN$ is renormalized in the medium.
The incoming $K^+$ has to produce a kaon, a free pion and a nuclear
excitation, so it is clear why the kinetic energy of the $K^+$ must
be larger than the pion mass as shown in fig.~{5}.

Looking again at fig.~{7} one realizes that not only one pion, but
also the other pion and also both pions simultaneously must be
modified by the nuclear medium. (The same applies to the intermediate
kaon, but we will see that the kaon modification in the medium is
negligible as compared to pion modification).

The imaginary part of the kaon selfenergy due to the pion cloud is
given by the diagram of fig.~{8}, except that its linear part in
density is to be subtracted to eliminate selfenergy parts which are
already included in the IA. Let $\tilde \Pi(k)$ be the $K^+$
selfenergy of the diagram of fig.~{8}, then the pionic cloud or MEC
contribution to the $K^+$ selfenergy, $\Pi^{\rm MEC}$, is given by:
\begin{eqnarray}
{\rm Im}\Pi^{\rm MEC}(k)
& = & {\rm Im}\tilde\Pi(k)
- \rho \left (
{\partial {\rm Im}\tilde\Pi(k)\over \partial \rho} \right
)_{\rho=0}
- \left (  {\rm Im}\tilde\Pi(k) \right
)_{\rho=0}
 , \label{Ee20}
\end{eqnarray}
but $\left (  {\rm Im}\tilde\Pi(k) \right )_{\rho=0} =0$ if the kaon
is on-shell because a free kaon is stable under strong interactions.

The $K^+$ selfenergy associated to the diagram of fig.~{8} is given
by:
\begin{eqnarray}
\tilde\Pi(k) &=& -{1\over 2}\int {d^4 q \over (2\pi)^4}
\int {d^4 q^\prime \over (2\pi)^4} D(q^\prime) D(q) D_K(k^\prime)
\nonumber\\
& \times & \sum_{ijl}
t_{K^+\pi^i\to K^l\pi^j}(k^\prime,q^\prime;k,-q)
t_{K^l\pi^j\to K^+\pi^i}(k,-q;k^\prime,q^\prime)\Big
|_{k^\prime=k-q-q^\prime}
\ . \label{Ee21}
\end{eqnarray}
Its imaginary part is easily evaluated by means of Cutkosky
rules \cite{ref17}: in all intermediate states cut by the dotted line
substitute:
\begin{eqnarray}
\Pi (k) &\rightarrow & 2 i \theta(k^0)
{\rm Im}~\Pi (k) \nonumber \\
D (q) &\rightarrow & 2 i \theta (q^{0}) {\rm Im} D(q) \nonumber \\
D_{K} (q) &\rightarrow & 2 i \theta (q^{0})
{\rm Im}~ D_{K} (q)
 \label{Ee22}
\end{eqnarray}
and put complex conjugate the amplitude above the dotted line.
Thus, by taking the imaginary part corresponding to cutting the three
meson propagators as shown by the horizontal line of fig.~{8}, one
obtains:
\begin{eqnarray}
{\rm Im}\tilde\Pi(k) &=& 2\int {d^4 q \over (2\pi)^4}
\int {d^4 q^\prime \over (2\pi)^4 }
\theta({q^0}^\prime)\theta(q^0)\theta({k^0}^\prime)
{\rm Im}D(q^\prime) {\rm Im}D(q)
{\rm Im}D_K(k^\prime)
\nonumber\\
& \times & \sum_\alpha |t^\alpha_{K\pi}(k^\prime,q^\prime;k,-q)|^2
\Big |_{k^\prime=k-q-q^\prime}
\ ,\label{Ee23}
\end{eqnarray}
where $\alpha$ runs over the indices $i,j,l$ in
$K^+\pi^i\to K^l\pi^j$.

By expanding the pion propagators above in powers of the density,
$D=D_0+D_{(1)}+\delta D$ (being $D_{(1)}$ of order $\rho$, and
$\delta D$ the remaining terms of higher orders), and using the
symmetry $q\leftrightarrow q^\prime$ (which holds due to the crossing
symmetry of the amplitudes), we obtain:
\begin{eqnarray}
\lefteqn{{\rm Im}\tilde\Pi(k) = 2\int {d^4 q \over (2\pi)^4}
\int {d^4 q^\prime \over (2\pi)^4 }
\theta(q^0)\theta({q^0}^\prime)\theta({k^0}^\prime){\rm
Im}D_K(k^\prime)}&&
\label{Ee24} \\
 &&\times
\sum_\alpha |t^\alpha_{K\pi}(k^\prime,q^\prime;k,-q)|^2
\Big |_{k^\prime=k-q-q^\prime}
\ \{  {\rm Im}D_0(q^\prime) {\rm Im}[D_0(q)+2 D_{(1)}(q)+
2\delta D(q) ] \nonumber \\
&& \phantom{\times\sum_\alpha
|t^\alpha_{K\pi}(k,-q;k^\prime,q^\prime)|^2
\Big |_{k^\prime=k-q-q^\prime}}
\ + {\rm Im}D_{(1)}(q^\prime){\rm Im}D_{(1)}(q)
 + {\cal O}(\rho^3)\}. \phantom{2\delta D(q) ]}
\nonumber
\end{eqnarray}
where the terms in $\delta D~D_{(1)}$ and $\delta D~\delta D$, of
at least order $\rho^3$, have been neglected. The term with ${\rm
Im}D_0(q'){\rm Im}D_0(q)$  and the term with $2{\rm Im}D_0(q'){\rm
Im}D_{(1)}(q)$ have to be subtracted because they are of zeroth and
first order in density, respectively; and the terms with $2{\rm
Im}D_0(q'){\rm Im}\delta D(q)$ and ${\rm Im}D_{(1)}(q^\prime){\rm
Im}D_{(1)}(q)$ must be kept since they are quadratic in density.

By subtracting the terms constant and linear in $\rho$  we obtain the
contributions to  Im$\Pi^{\rm MEC}$. Those contributions are diagrams
$d1$, $d2$, $d3$ and $d4$ depicted in fig.~{9}, and their explicit
expressions are given by:

\begin{eqnarray}
\lefteqn{{\rm Im}\Pi^{\rm MEC}(k) = 2\int {d^4 q \over (2\pi)^4}
\int {d^4 q^\prime \over (2\pi)^4 }
\theta({q^0}^\prime)\theta(q^0)\theta({k^0}^\prime){\rm
Im}D_K(k^\prime)}
\label{Ee25}\\
&\times
 \sum_\alpha  |t^\alpha_{K\pi}(k^\prime,q^\prime;k,-q)|^2
\Big |_{k^\prime=k-q-q^\prime}
\{ 2{\rm Im}D_0(q^\prime) {\rm Im}\delta D(q)\  &(d1)
\nonumber \\
\ &
\phantom{\times
\sum_\alpha  |t^\alpha_{K\pi}(k^\prime,q^\prime;k,-q)|^2
\Big |_{k^\prime=k-q-q^\prime}}
+
{\rm Im}D_{(1)}(q^\prime){\rm Im}D_{(1)}(q)\} &(d2+d3+d4)
\ . \nonumber
\end{eqnarray} %
Now we consider the first term, diagram $d1$. If we look at the
diagram of fig.~{6}b we can write the amplitude for this process
following the same rules used so far and by means of Cutkosky rules,
we obtain:
\begin{eqnarray}
&&3{\rm Im}\tilde t(s')=
\label{Ee26}\\
\nonumber
&&-2\int {d^4 q'\over (2\pi)^4}
                          \sum_\alpha |t^\alpha_{K\pi}(k',q';k,p)|^2
\theta({q^0}')\theta({k^0}')
{\rm Im} D_0(q'){\rm Im} D_K(k')\Big |_{k^\prime=k+p-q^\prime},
\end{eqnarray}
with $s'=(k+p)^2$. If we substitute this in eq.~(\ref{Ee25}) we
find
\begin{eqnarray}
{\rm Im}\Pi^{\rm d1}(k)
=-6\int {d^3 q\over (2\pi)^3}\int_0^\infty{dq^0\over 2\pi}
                                {\rm Im}\tilde t(u)~{\rm Im}
\delta D(q)
= {\rm Im}\delta\Pi(k)\ \ ,\label{Ee27}
\end{eqnarray}
which is the same result as eq.~(\ref{Ee19}). At the first sight the
upper limit in the $q^0$ integration in eq.~(\ref{Ee26}) is different
than in eq.~(\ref{Ee19}), but the condition Im$\tilde t(u)\neq 0$
makes $q^0$ smaller than $k^0-E(x_0)$, and we regain the same limit.
It is interesting to note that in spite of renormalizing the only
existing pion line in eq.~(\ref{Ee19}) one obtains the same result as
here where we have renormalized either of the two pions in
fig.~{7b}. The reason is that the use of a crossing symmetric
amplitude and the equal contribution of the terms Im$\tilde t(s)$ and
Im$\tilde t(u)$ of eq.~(\ref{e15}) in eq.~(\ref{Ee19}) accounts for
that. This can be better visualized if we make use of a model
$K^+\pi^-$ scattering consisting of a resonant $K^*$ pole (as in  the
$p$-wave amplitude of the model we use). If in fig.~{6}b and
fig.~{6}c we renormalize and fold the external pion we obtain two
diagrams like in fig.~{7b} where in one case one pion is
renormalized and in the other case the other pion is renormalized.
Hence we conclude that Im$\Pi^{d1}(k)$ is exactly the same
contribution obtained before in eq.~(\ref{Ee19}). However, we get now
new contributions from the terms $d2$, $d3$, $d4$ which renormalize
the two pions simultaneously.

The second term in ${\rm Im}\Pi^{\rm MEC}$ comes from ${\rm
Im}D_{(1)}{\rm Im}D_{(1)}$. It contains new reaction channels,
namely,  the $K^+$ decaying into $K~ph~ph$, $K~ph~\Delta h$ and
$K~\Delta h~\Delta h$. These have not been considered before. To
evaluate them we need $\sum_\alpha |t_{K\pi}^\alpha(k,-q;k',q')|^2$
with the $t-$matrix off-shell for both $q$ and $q'$. Given the small
contribution from the $p-$wave part, the only one with an angular
dependence in ${\bf q}^\prime$, we substitute $|t_{K\pi}|^2$ by an
angular average in eq.~(\ref{Ee26}), and hence we find for
$u>(m_\pi+m_K)^2$
\begin{eqnarray}
\sum_\alpha
|t_{K\pi}^\alpha(k',q';k,-q)|^2_{\rm ave}&=&
{-3 {\rm Im} \tilde t(u)
\over 2\int {d^4 q^\prime \over (2\pi)^4 }
\theta({k^0}^\prime){\rm Im}D_K(k^\prime)\theta({q^0}^\prime)
{\rm Im}D_0({q}^\prime)
\Big |_{k^\prime=k-q-q^\prime}}\nonumber \\
\label{Ee28}
\end{eqnarray}
which after the evaluation of the denominator gives
\begin{equation}
\sum_\alpha
|t_{K\pi}^\alpha(k',q';k,-q)|^2_{\rm ave}
=-{8\pi \sqrt{u}\over q_{\rm CM}}3{\rm Im}\tilde t(u)\simeq
 f(u).
\label{Ee29}
\end{equation}
with $u=(k-q)^2$ and $q_{\rm CM}$ the $K^+\pi$ CM momentum for $K^+$
and $\pi$ on-shell. The function $f(u)$ for any value of $u$, in the
model which we use, is explicitly written in appendix B, where some
approximations are made which are consistent with the model itself.
Now we use $ f(u)$ to calculate the new channels. By expanding the
first medium correction to the pion propagator into $ph$ and $\Delta
h$ components as  $D_{(1)}=D_{(1)}^{ph}+D_{(1)}^{\Delta h}$, one
obtains:
\begin{equation}\label{Ee30}
\begin{array}{lllr}
\multicolumn{3}{l} {{\rm Im}\Pi^{\rm MEC}(k)} & \\
&=&\ \ {\rm Im}\delta\Pi(k) +& (d1) \\
 & \multicolumn{2}{l} {+2\int{d^4 q\over(2\pi)^4}
\int{d^4 q'\over(2\pi)^4}
\theta(q^0)\theta({q^0}')\theta({k^0}')
{\rm Im}D_K(k') f((k-q)^2)\ \ \ } & \\
&&\times \{ {\rm Im}D_{(1)}^{ph}(q')~{\rm Im}D_{(1)}^{ph}(q) +
 & (d2) \\
&&\ + {\rm Im}D_{(1)}^{ph}(q')~{\rm Im}D_{(1)}^{\Delta h}(q)
    + {\rm Im}D_{(1)}^{ph}(q)~{\rm Im}D_{(1)}^{\Delta h}(q') +
 & (d3) \\
&&\ + {\rm Im}D_{(1)}^{\Delta h}(q')~{\rm Im}D_{(1)}^{\Delta h}(q)
\} & (d4) \\
\end{array}
\end{equation}
Diagrams $d2$, $d3$ and $d4$ are
genuine new channels and correspond, respectively, to the processes:
$K~\to~K~ph~ph$,
$K~\to~K~ph~\Delta h$ and $K~\to~K~\Delta h~\Delta h$.
Those processes have the following thresholds:
$T_K\geq 222~{\rm MeV}>m_\pi$ for diagram $d1$,
$T_K>0    $ for diagram $d2$,
$T_K\geq 181~{\rm MeV}>m_\pi$ for diagram $d3$ and
$T_K\geq 392~{\rm MeV}>2 m_\pi$ for diagram $d4$.

We do not evaluate the contribution of diagram $d4$ because its
threshold is close to the highest energies we consider and we expect
it to be small. We have done the calculations up to second order in
density for the processes$K\to K~ph~ph$. Higher orders have been
considered for $K\to K~\pi~ph$ , $K\to K~\pi~\Delta h$ and $K\to
K~ph~\Delta h$. But despite that, we have found that their behavior
is quadratic in density, so higher order corrections are negligible.
In fig.~{10}, results for Im$\Pi^{\rm d1}$ and Im$\Pi^{\rm d3}$ with
$T_K=450$~MeV are depicted with crosses for different densities, the
lines shown for comparison are exact quadratic functions.

On the other hand in Im$\Pi^{d1}$ so far we have considered only the
second order terms coming from the iterated $ph$ or $\Delta h$
excitations like in fig.~{2}d. Now it is easy to include the
diagrams of the same order in the density  corresponding to a
simultaneous excitation of two particles-two holes  by the pion
(related to the second order pion proper selfenergy), this
contribution is given by diagram $d5$, shown in fig.~{11}. Its
contribution to the imaginary part of the $K^+$-selfenergy is given
by the same expression than Im$\Pi^{\rm d1}$ of eq.~(\ref{Ee27})
where instead of Im$\delta D$ we are considering Im$\delta D^{\rm
d5}$, which is the modification of the pion propagator due to the
$2p2h$ channel of pion absorption:
\begin{eqnarray}
{\rm Im}\Pi^{\rm d5}(k)
&=&-6\int {d^3 q\over (2\pi)^3}\int_0^\infty{dq^0\over 2\pi}
                                {\rm Im}\tilde t(u)~{\rm
Im}\delta D^{\rm d5}(q)\ \ ,\label{Ee31}\\
{\rm Im}\delta D^{\rm d5}(q)&=&D_0^2(q)~{\rm Im}\Pi_\pi^{2p2h}(q)
\ ,
\nonumber
\end{eqnarray}
where Im$\Pi_\pi^{2p2h}$ is the pion selfenergy due to the
$2p2h$ channel of pion absorption.
For
this we take the model of \cite{ref16}, which in lowest order in
density contains the same input as here, but
we have simplified it and rewriten the second order part of it
as in ref.~\cite{ref22} which
gives rise to about the same results for pionic atoms. Since
this selfenergy is for pions on-shell, we modify it by
multiplying by the ratio of phase space for $2p2h$ excitation
for the off-shell and on-shell situations and by the
pion-nucleon squared form-factor $F^2(q)$ (given in appendix A)

\begin{eqnarray}
{\rm Im}\Pi_\pi^{2p2h}(q)&=&-4\pi{{\bf q}}^2~{\rm
Im}C_0~\rho^2~F^2(q){phase (q^0,{\bf q})\over phase(m_\pi,
{\bf 0})}\ ,\nonumber\\
{phase (q^0,{\bf q})\over phase(m_\pi,
{\bf 0})}&=&\theta(4Mq^0-{{\bf q}}^2)\sqrt{{4Mq^0-{{\bf q}}^2\over
4Mm_\pi}}
+{\cal O}(k_F)\ ,\label{Ee32}\\
{\rm Im}C_0 &=& 0.096 m_\pi^{-6}\ \nonumber ,
\end{eqnarray}
where the phase-space ratio has been taken at $\rho=0$. Only the
$p-$wave part of the pionic optical potential has been written, since
the $s-$wave part contribution is much smaller than this for the
relevant values of ${\bf q}$ involved.

One could also think about effects from the renormalization of the
intermediate kaon propagator. The impulse approximation $t_{KN}~\rho$
from Appendix C  provides the dominant part of the kaon selfenergy.
{}From fig.~{5} one can see that $-{\rm Im}\Pi (k) / m_{K}^{2}\sim
0.04$ is much smaller than $-{\rm Im}\Pi_{\pi} (k) / {m_\pi}^{2}$ and
thus the corrections from this source can be estimated reasonably
smaller than those obtained from pion renormalization.

The contributions of all MEC diagrams to the imaginary part of the
$K^+$ selfenergy  are  approximately quadratic in density. In
fig.~{12} the contribution of each of the diagrams $d1$, $d2$, $d3$
and $d5$ at density $\rho_0$ are shown for different kinetic energies
of the kaon. Also the total of the MEC contributions, $d1+d2+d3+d5$,
is shown. For comparison the exact result for the ``standard''
calculation, Im$\delta \Pi$=Im$\Pi^{d1}$, is depicted with dashed
line. By itself it is much smaller than the total MEC contributions
coming from diagrams $d1$, $d2$, $d3$ plus $d5$. The diagram $d2$ is
the most important MEC correction for low energies, but for higher
energies the most relevant is $d3$, being more than half the total
MEC effect for $T_K=450$~MeV. The contribution of diagram $d5$ is
negligible as seen in figure, if we had considered the Im$C_0$
parameter of the pionic atoms optical potential of eq.~(\ref{Ee32})
to be up to four times larger, such as it is in certain
parametrizations found in the literature, its contribution would
still be negligible as compare to the total MEC result. In fig.~{5}
we display the total MEC value together with the value of the IA. We
see that the contribution due to the pionic cloud is sizable in
relation to the dominant term which comes from the IA. We have
checked that these curves are fairly stable under a reasonable
modification of the LLEE parameter $g^\prime$.

The selfenergy  $\Pi^{\rm MEC}({\bf r})$ in finite nuclei is obtained
by substituting $\rho$ by $\rho({\bf r})$ in the nuclear matter
results. This is shown in ref.~\cite{ref16} to be practical and
accurate for the $s-$wave part, which gives practically the whole
contribution here.

The results shown in this section and in section 6 complement and
correct our preliminary results exposed in ref.~\cite{ref23}

We have observed that the resonant
part of the $t_{K\pi}$-matrix does not contribute significantly to
the imaginary part of the kaon-nucleus optical potential. Its
contribution is smaller than one per cent for $T_K\leq 550~$MeV/$c$.
In other words Im$\Pi^{\rm MEC}$ is, in very good approximation,
proportional to the parameter $\beta'_0$. This parameter is
given in the work of \cite{12} by
$$ \beta'_0 = -{8\pi\over 3}
(m_\pi+m_K) (a_1^2+2 a_3^2)\,, $$
where $a_{2I}$ is the scattering length in the isospin $I$ channel.

Then we can consider a
simplified model, which consists of taking a purely constant value for
$|t_{K\pi}|^2$, with $\beta'_0$ given by
the scattering lengths. In
particular the quantity $f(u)$ (eq.~(\ref{Ee53})), relevant for
Im$\Pi^{\rm MEC}$, is now a constant. In the model of ref.~\cite{12}
$\beta'_0$ is constrained by on-shell data at threshold which are
taken from the analysis of ref.~\cite{ref25}. This simplified model
saves a lot of computing time because certain integrals in
eq.~(\ref{Ee30}) can be done trivially due to the constancy of $f(u)$.
\vspace{1cm}

\noindent
{\large \bf 5.- Real part of the $K^+$ optical potential}

\nobreak
Now, let us pay attention to the calculation of the real part of the
$K^+$ selfenergy due to the pionic cloud in the nucleus.

We go back to eq.~(\ref{Ee12}) for $\delta\Pi(k)$ and substitute the
isoscalar averaged $t$-matrix of eq.~(\ref{e15}) with its dispersion
relation of eq.~(\ref{e16}). Then, we consider separately the
contributions to $\delta\Pi$ coming from the analytical part of
$\tilde t$ (this is $P$ ) and from its dispersive part (related to
Im$\tilde t$):
\begin{eqnarray}
\delta\Pi&=&{3}~i \int \frac{d^{4} q}{(2 \pi)^{4}}
\delta D(q)~ \left( P(s)+
\int_{x_0}^\infty {dx\over \pi}{{\rm Im}\tilde t(x)\over
x_0-x}  {s-x_0\over s-x+i\epsilon}\right ) \  , \label{Ee33}
\end{eqnarray}
where use has been made of crossing symmetry to cancel the factor
1/2. As $P(s)$ is a real polynomial in $q^0$ and, by doing a Wick
rotation for $q^0$ as depicted in fig.~{4}, it can be proved that the
first part (that going with $P(s)$) is real. By doing the same Wick
rotation, one can see that the second part (going with ${\rm
Im}\tilde t(x)$) is complex, its imaginary part being given by
eq.~(\ref{Ee19}). This second part is  linear in Im$\tilde t(x)$. Due
to the optical theorem Im$\tilde t(x)\propto | \tilde t( x)|^2$. So,
the first part is of order $\tilde t$ and the second one is of order
$\tilde t^2$. We are keeping the leading order contribution to both,
${\rm Re}\delta\Pi$ and ${\rm Im}\delta\Pi$, this is: order $\tilde
t$ for ${\rm Re}\delta\Pi$ and order $\tilde t^2$ for ${\rm
Im}\delta\Pi$ . Within the same approximation , one should neglect
for the real part the contribution of the diagrams $d2$, $d3$ and
$d5$. Note that to order $|\tilde t|^2$ there would be more diagrams
besides these, which do not contribute, or contribute little, to the
imaginary part, for instance $d4$.

With this approach of keeping the dominant order in $\tilde t$ and
considering the incoming $K^+$ on-shell:
\begin{eqnarray}{\rm Re}\delta\Pi(k)&=&
-{3}\int {d^4 q\over (2\pi)^4}{\rm
Im}\delta D(q)~P(s)\label{Ee34}\\
\label{Ee35}
&=&-3\int {d^3{\bf q}\over (2\pi)^3}\int_0^\infty {dq^0\over \pi} {\rm Im}
\delta D(q)
[{\alpha_0\over 2}+\beta_0(m_K^2+{q^0} ^2-{\bf q}^2)].
\end{eqnarray}
\noindent
Where we have used
\begin{eqnarray}
P(x) &=& {\alpha_0\over 2}~+~\beta_0~x \ ,\label{Ee36}
\end{eqnarray}
which, in the model of \cite{12}, is the dominant contribution to
$P(x)$ and comes from the $s-$wave, the $p-$wave contribution has
been neglected. $\alpha_0$, $\beta_0$ are given in table 1. In this
approximation Re$\delta\Pi$ is independent ot the $K^+$ kinetic
energy. Eq.~(\ref{Ee35}) is the estimation we are going to use for
the real part of the $K^+$-selfenergy. Observe that for Re$\delta\Pi$
we have the pion four-momentum without any phase-space restriction.
But also notice that the relevant results are coming from the
$q$-values such that ${\rm Im}\delta D(q)$ is large, and this happens
for small values of ${\bf q}$ and $q^0$, because in the limit of $q$
large (${\bf q}\to\infty$ or $q^0\to\infty$) the pion selfenergy,
which makes $D(q)$ different to $D_0(q)$, goes to zero. For the
purpose of evaluating ${\rm Re}\delta\Pi$, we split it in different
parts as follows:
\break
\begin{eqnarray}
{\rm
Re}\delta\Pi =& \alpha_0~\gamma_1~+~2\beta_0~(\gamma_2~+~\gamma_3)
,\label{Ee37}\\
\gamma_1 =&
-{3\over 2}\int
{d^3{\bf q}\over (2\pi)^3}\int_0^\infty {dq^0\over \pi} {\rm
Im}\delta D(q) \phantom{(m_K^2-{\bf q}^2)}
 \equiv & \int{d^3{\bf q}\over (2\pi)^3}{\delta N({\bf q})\over
2\omega({\bf q})}
,\nonumber\\
\gamma_2 =&
-{3\over 2}\int {d^3{\bf q}
\over (2\pi)^3}\int_0^\infty {dq^0\over \pi} {\rm
Im}\delta D(q)
(m_K^2-{\bf q}^2)
\equiv & \int{d^3{\bf q}\over (2\pi)^3}{\delta N({\bf q})\over
2\omega({\bf q})}
(m_K^2-{\bf q}^2)\ ,\nonumber\\
\gamma_3 =&
-{3\over 2}\int {d^3{\bf q}
\over (2\pi)^3}\int_0^\infty {dq^0\over \pi} {\rm
Im}\delta D(q)
 {q^0}^2 \phantom{(m_K^2-)}.&
\nonumber
\end{eqnarray}
\noindent
By doing the numerical evaluation using the pion selfenergy
of Appendix A, the results are:
$\gamma_1=0.02$~fm$^{-2}$,
$\gamma_2=-0.04$~fm$^{-4}$,
$\gamma_3=-0.006$~fm$^{-4}$ evaluated at
$\rho=\rho_0$.
Notice that ${\rm Re}\delta\Pi$
is independent of the energy of the $K^+$, then
\begin{equation}
{\rm
Re}\delta\Pi(k;{\bf r})=
{\rm Re}\delta\Pi ({\bf r})={\rm Re}\delta\Pi (\rho=\rho_0)\left
({\rho({\bf r})\over \rho_0}\right )^2 \equiv {\rm Re}B\left
({\rho({\bf r})\over \rho_0}\right )^2.
\label{A38}
\end{equation}
For the parametrizations I and III of ref.~\cite{12}, we obtain the
values of ${\rm Re}\delta\Pi (\rho=\rho_0)\equiv {\rm Re}B$
shown in table~1 and
fig.~{13}.  ${\rm Re}\delta\Pi_{\rm stat} $ has been obtained by
taking the static approximation $q^0=0$ in eq.~(\ref{Ee35}), which
amounts to taking $\gamma_3=0$ in eq.~(\ref{Ee37}). Noting that
$\gamma_3/\gamma_2\simeq 0.15$, we see that this static approximation
is not bad in this case. We see that different off-shell
extrapolations provide very different results for the real part of
the MEC selfenergy of kaons, being possible to obtain different signs
for it. ${\rm Re}\delta\Pi >0$ for parametrization I,  and  ${\rm
Re}\delta\Pi <0$ for parametrization III. Parametrization IV is
obtained by imposing Re$\delta\Pi$=0, which gives
$\alpha_0=-2.8,~\beta_0=-0.61$fm$^2$. Note that $\alpha_0,~\beta_0$
are related in order to reproduce the scattering lengths, see
Appendix B.

In fig.~{13} the real parts calculated at order $\tilde t$ are
displayed with labels I, III and IV, depending on the parametrization
used. The result I is of the same order of the impulse approximation
IA, but III is smaller and  has different sign. We also present the
result of doing the calculation in the spirit of ref.~\cite{12}, that
is, in the static approximation and keeping the whole Re$\tilde t$
(including $P(x)$ and the part related to Im$\tilde t$), in this case
we use parametrization I, the result is labeled I$_{\rm JK}$. We see
that I$_{\rm JK}$ is quite different from I, this means than the term
coming from Im$\tilde t$ is not small, but we should not calculate
one of those contributions but all of them which are of the same
order in $\tilde t$.

Given the large sensitivity of the real part to the uncertainties of
the off-shell extrapolation for the $\tilde t$-matrix, we think that
the computation of the real part is presently beyond the scope of the
microscopical approach.

\vspace{1cm}

\noindent
{\large \bf 6.- Results: K$^+$-nucleus cross section}

\nobreak

In fig.~{14} we show the results for $d\sigma\over d\Omega$ for
$K^+$ with energy $T_K=450~$MeV scattered by $^{12}$C. The result
with the impulse approximation is compared to those including pion
cloud effects, using parametrizations I and III, the result using
parametrization IV, not shown,( Re($\Pi^{\rm MEC})=0$) is in between.
We find that the
inclusion of MEC effects provides some improvement in comparison with
the experimental data. In fig.~{15} the same is compared for the
$^{40}$Ca nucleus. The effect of pion cloud is very similar to the
case of $^{12}$C. In both cases the Coulomb interaction is neglected.
Here we show the results in order to
see the size and shape of the MEC effects in the
differential cross-section.
Some other theoretical corrections, as discussed later, should also be
included for a proper comparison with experiments.

The total cross-section of $K^+$ scattered by $^{12}$C versus
kinetical energy of the $K^+$ is shown in fig.~{16a} . The
experimental data with a cross are from ref.~\cite{8}, the data with
a diamond are from ref.~\cite{4}, only the statistical errors are
included, the systematic errors, not shown, are larger.
The dashed line labeled IA corresponds to
the impulse approximation. The dotted and solid
lines include MEC effects using
parametrizations I, III and IV. Note that Im$\Pi^{\rm MEC}$, as
computed in section 3 does not depend on the parametrization.

For low energies the resulting cross-section including MEC depends a
lot on the real part of the optical potential and hence on the
parametrization used. But it is less dependent for higher energies.
Given that the real part is not at all under control from the model,
we take hereafter as a reference the line labeled IV, which amounts
to neglecting the MEC effects for the real part of the
optical potential. Figure 16a shows that the inclusion of MEC effects
in the imaginary part significantly improves on the impulse
approximation, bringing the cross-section closer to the experimental
one, and showing that MEC effects are large enough to have to be
considered in this process.

We have used Arndt's phase-shifts \cite{ref27}.
The calculated $K^+$ nucleus total cross section would have been larger
if we had used Martin's \cite{ref26}
phase shifts rather than Arndt's \cite{ref27} for the $KN$ scattering
amplitudes as shown in ref.~\cite{2}. The analysis of \cite{ref27} is
more recent than the one in \cite{ref26}. On the other hand a recent
reanalysis \cite{ref28} of the $KN$ data seems to favor Martin's
phase shifts.
%

These theoretical
uncertainties in the $K^+$-nucleus cross-sections partially
cancel if a quotient of cross-sections is taken.
Same is true for systematic
errors in the experimental measurements of the total $K^+$-nucleus
cross-sections. Then, the magnitude usually calculated and compared
with the experiment in most of research works is the ratio over the
$K^+$ deuterium total cross-section:
$R[A]={\sigma(K^+~A)\over (A/2)\sigma(K^+~^{2}H)}$ for a nucleus of
mass number $A$.
In particular for $^{12}C$ the
following ratio $R$ is defined:
\begin{equation}
R={\sigma(K^+~^{12}C)\over 6\sigma(K^+~^{2}H)}\ ,
\label{A39}
\end{equation}
where the factor 6 in the denominator is included to emphasize the
closeness of the ratio to unity.

One should notice that for the magnitude $R$, being a quotient of
cross-sections, the possible uncertainties due to the use of different
phase-shifts partially cancel and they are not relevant. So we present
in fig.~16b the same results as in fig.~16a but for the ratio $R$,
which has less error than the total cross section.
The dashed line correspond to the IA calculation, and the other lines
include IA+MEC, using for MEC effects the parametrizations I,III and
IV as labeled.

So far we have calculated MEC effects on the total $K^+$-nucleus
cross-section for $^{12}C$. We have shown that these effects are large
enough as compared to the IA approximation and, then, they need to be
considered.

But, for doing a meaningful comparison of theoretical calculations
with experiments, a more realistic $K^+$-nucleus standard optical
potential (than the crude $t\rho$ impulse approximation used in
fig.~14-16)
should be considered.
There are corrections over this $t\rho$ which should be
included like off-shell range, binding energy,... Also, there are
nucleon-nucleon correlations which contribute to the optical potential
in second order. These correlation effects are approximately $\rho^2$
by nature, which is the same form as the MEC contributions to the
optical potential.

Both kinds of corrections to the optical potential (and their effect
on the $K^+$-nucleus cross-sections) have been calculated and/or
estimated in the works of Siegel, Kaufmann and Gibbs \cite{1,2}.
As result of their studies, they provide a band of uncertainty for the
conventional calculation, the boundaries of the band are determined by
varying the parameters in the theoretical model.

We show in fig.~17 with dashed-lines the band of theoretical
calculations for $R$ with the conventional microscopic optical
potential of ref.~\cite{2}, for $p<500$~MeV/$c$ the results quoted
are from ref.~\cite{6}.
%
\begin{figure}
  \begin{center}
\setlength{\unitlength}{0.240900pt}
\ifx\plotpoint\undefined\newsavebox{\plotpoint}\fi
\sbox{\plotpoint}{\rule[-0.200pt]{0.400pt}{0.400pt}}%
\begin{picture}(1500,900)(0,0)
\font\gnuplot=cmr10 at 10pt
\gnuplot
\sbox{\plotpoint}{\rule[-0.200pt]{0.400pt}{0.400pt}}%
\put(220.0,113.0){\rule[-0.200pt]{4.818pt}{0.400pt}}
\put(198,113){\makebox(0,0)[r]{0.8}}
\put(1416.0,113.0){\rule[-0.200pt]{4.818pt}{0.400pt}}
\put(220.0,203.0){\rule[-0.200pt]{4.818pt}{0.400pt}}
\put(198,203){\makebox(0,0)[r]{0.85}}
\put(1416.0,203.0){\rule[-0.200pt]{4.818pt}{0.400pt}}
\put(220.0,293.0){\rule[-0.200pt]{4.818pt}{0.400pt}}
\put(198,293){\makebox(0,0)[r]{0.9}}
\put(1416.0,293.0){\rule[-0.200pt]{4.818pt}{0.400pt}}
\put(220.0,383.0){\rule[-0.200pt]{4.818pt}{0.400pt}}
\put(198,383){\makebox(0,0)[r]{0.95}}
\put(1416.0,383.0){\rule[-0.200pt]{4.818pt}{0.400pt}}
\put(220.0,473.0){\rule[-0.200pt]{4.818pt}{0.400pt}}
\put(198,473){\makebox(0,0)[r]{1}}
\put(1416.0,473.0){\rule[-0.200pt]{4.818pt}{0.400pt}}
\put(220.0,562.0){\rule[-0.200pt]{4.818pt}{0.400pt}}
\put(198,562){\makebox(0,0)[r]{1.05}}
\put(1416.0,562.0){\rule[-0.200pt]{4.818pt}{0.400pt}}
\put(220.0,652.0){\rule[-0.200pt]{4.818pt}{0.400pt}}
\put(198,652){\makebox(0,0)[r]{1.1}}
\put(1416.0,652.0){\rule[-0.200pt]{4.818pt}{0.400pt}}
\put(220.0,742.0){\rule[-0.200pt]{4.818pt}{0.400pt}}
\put(198,742){\makebox(0,0)[r]{1.15}}
\put(1416.0,742.0){\rule[-0.200pt]{4.818pt}{0.400pt}}
\put(220.0,832.0){\rule[-0.200pt]{4.818pt}{0.400pt}}
\put(198,832){\makebox(0,0)[r]{1.2}}
\put(1416.0,832.0){\rule[-0.200pt]{4.818pt}{0.400pt}}
\put(220.0,113.0){\rule[-0.200pt]{0.400pt}{4.818pt}}
\put(220,68){\makebox(0,0){400}}
\put(220.0,812.0){\rule[-0.200pt]{0.400pt}{4.818pt}}
\put(423.0,113.0){\rule[-0.200pt]{0.400pt}{4.818pt}}
\put(423,68){\makebox(0,0){500}}
\put(423.0,812.0){\rule[-0.200pt]{0.400pt}{4.818pt}}
\put(625.0,113.0){\rule[-0.200pt]{0.400pt}{4.818pt}}
\put(625,68){\makebox(0,0){600}}
\put(625.0,812.0){\rule[-0.200pt]{0.400pt}{4.818pt}}
\put(828.0,113.0){\rule[-0.200pt]{0.400pt}{4.818pt}}
\put(828,68){\makebox(0,0){700}}
\put(828.0,812.0){\rule[-0.200pt]{0.400pt}{4.818pt}}
\put(1031.0,113.0){\rule[-0.200pt]{0.400pt}{4.818pt}}
\put(1031,68){\makebox(0,0){800}}
\put(1031.0,812.0){\rule[-0.200pt]{0.400pt}{4.818pt}}
\put(1233.0,113.0){\rule[-0.200pt]{0.400pt}{4.818pt}}
\put(1233,68){\makebox(0,0){900}}
\put(1233.0,812.0){\rule[-0.200pt]{0.400pt}{4.818pt}}
\put(1436.0,113.0){\rule[-0.200pt]{0.400pt}{4.818pt}}
\put(1436,68){\makebox(0,0){1000}}
\put(1436.0,812.0){\rule[-0.200pt]{0.400pt}{4.818pt}}
\put(220.0,113.0){\rule[-0.200pt]{292.934pt}{0.400pt}}
\put(1436.0,113.0){\rule[-0.200pt]{0.400pt}{173.207pt}}
\put(220.0,832.0){\rule[-0.200pt]{292.934pt}{0.400pt}}
\put(45,472){\makebox(0,0){\shortstack{ $R=$\\${\sigma_{\rm tot}[^{12}C]\over
6\sigma_{\rm tot}[^2H]}$}}}
\put(828,23){\makebox(0,0){$p_{\rm lab}[{\rm MeV}/c]$}}
\put(17548,922){\makebox(0,0){MEC effects }}
\put(220.0,113.0){\rule[-0.200pt]{0.400pt}{173.207pt}}
\put(1306,767){\makebox(0,0)[r]{Krauss}}
\put(1350,767){\makebox(0,0){$\times$}}
\put(327,715){\makebox(0,0){$\times$}}
\put(398,602){\makebox(0,0){$\times$}}
\put(437,564){\makebox(0,0){$\times$}}
\put(485,557){\makebox(0,0){$\times$}}
\put(542,595){\makebox(0,0){$\times$}}
\put(708,564){\makebox(0,0){$\times$}}
\put(739,539){\makebox(0,0){$\times$}}
\put(779,564){\makebox(0,0){$\times$}}
\put(834,606){\makebox(0,0){$\times$}}
\put(856,508){\makebox(0,0){$\times$}}
\put(909,492){\makebox(0,0){$\times$}}
\put(1328.0,767.0){\rule[-0.200pt]{15.899pt}{0.400pt}}
\put(1328.0,757.0){\rule[-0.200pt]{0.400pt}{4.818pt}}
\put(1394.0,757.0){\rule[-0.200pt]{0.400pt}{4.818pt}}
\put(327.0,629.0){\rule[-0.200pt]{0.400pt}{41.435pt}}
\put(317.0,629.0){\rule[-0.200pt]{4.818pt}{0.400pt}}
\put(317.0,801.0){\rule[-0.200pt]{4.818pt}{0.400pt}}
\put(398.0,537.0){\rule[-0.200pt]{0.400pt}{31.317pt}}
\put(388.0,537.0){\rule[-0.200pt]{4.818pt}{0.400pt}}
\put(388.0,667.0){\rule[-0.200pt]{4.818pt}{0.400pt}}
\put(437.0,498.0){\rule[-0.200pt]{0.400pt}{32.040pt}}
\put(427.0,498.0){\rule[-0.200pt]{4.818pt}{0.400pt}}
\put(427.0,631.0){\rule[-0.200pt]{4.818pt}{0.400pt}}
\put(485.0,530.0){\rule[-0.200pt]{0.400pt}{13.009pt}}
\put(475.0,530.0){\rule[-0.200pt]{4.818pt}{0.400pt}}
\put(475.0,584.0){\rule[-0.200pt]{4.818pt}{0.400pt}}
\put(542.0,541.0){\rule[-0.200pt]{0.400pt}{26.017pt}}
\put(532.0,541.0){\rule[-0.200pt]{4.818pt}{0.400pt}}
\put(532.0,649.0){\rule[-0.200pt]{4.818pt}{0.400pt}}
\put(708.0,528.0){\rule[-0.200pt]{0.400pt}{17.345pt}}
\put(698.0,528.0){\rule[-0.200pt]{4.818pt}{0.400pt}}
\put(698.0,600.0){\rule[-0.200pt]{4.818pt}{0.400pt}}
\put(739.0,519.0){\rule[-0.200pt]{0.400pt}{9.636pt}}
\put(729.0,519.0){\rule[-0.200pt]{4.818pt}{0.400pt}}
\put(729.0,559.0){\rule[-0.200pt]{4.818pt}{0.400pt}}
\put(779.0,530.0){\rule[-0.200pt]{0.400pt}{16.381pt}}
\put(769.0,530.0){\rule[-0.200pt]{4.818pt}{0.400pt}}
\put(769.0,598.0){\rule[-0.200pt]{4.818pt}{0.400pt}}
\put(834.0,562.0){\rule[-0.200pt]{0.400pt}{20.958pt}}
\put(824.0,562.0){\rule[-0.200pt]{4.818pt}{0.400pt}}
\put(824.0,649.0){\rule[-0.200pt]{4.818pt}{0.400pt}}
\put(856.0,490.0){\rule[-0.200pt]{0.400pt}{8.672pt}}
\put(846.0,490.0){\rule[-0.200pt]{4.818pt}{0.400pt}}
\put(846.0,526.0){\rule[-0.200pt]{4.818pt}{0.400pt}}
\put(909.0,438.0){\rule[-0.200pt]{0.400pt}{26.017pt}}
\put(899.0,438.0){\rule[-0.200pt]{4.818pt}{0.400pt}}
\put(899.0,546.0){\rule[-0.200pt]{4.818pt}{0.400pt}}
\put(1306,722){\makebox(0,0)[r]{Mardor}}
\put(1350,722){\raisebox{-.8pt}{\makebox(0,0){$\Diamond$}}}
\put(323,753){\raisebox{-.8pt}{\makebox(0,0){$\Diamond$}}}
\put(434,557){\raisebox{-.8pt}{\makebox(0,0){$\Diamond$}}}
\put(542,512){\raisebox{-.8pt}{\makebox(0,0){$\Diamond$}}}
\put(707,541){\raisebox{-.8pt}{\makebox(0,0){$\Diamond$}}}
\put(832,557){\raisebox{-.8pt}{\makebox(0,0){$\Diamond$}}}
\put(902,510){\raisebox{-.8pt}{\makebox(0,0){$\Diamond$}}}
\put(1328.0,722.0){\rule[-0.200pt]{15.899pt}{0.400pt}}
\put(1328.0,712.0){\rule[-0.200pt]{0.400pt}{4.818pt}}
\put(1394.0,712.0){\rule[-0.200pt]{0.400pt}{4.818pt}}
\put(323.0,701.0){\rule[-0.200pt]{0.400pt}{25.054pt}}
\put(313.0,701.0){\rule[-0.200pt]{4.818pt}{0.400pt}}
\put(313.0,805.0){\rule[-0.200pt]{4.818pt}{0.400pt}}
\put(434.0,528.0){\rule[-0.200pt]{0.400pt}{13.972pt}}
\put(424.0,528.0){\rule[-0.200pt]{4.818pt}{0.400pt}}
\put(424.0,586.0){\rule[-0.200pt]{4.818pt}{0.400pt}}
\put(542.0,489.0){\rule[-0.200pt]{0.400pt}{11.081pt}}
\put(532.0,489.0){\rule[-0.200pt]{4.818pt}{0.400pt}}
\put(532.0,535.0){\rule[-0.200pt]{4.818pt}{0.400pt}}
\put(707.0,517.0){\rule[-0.200pt]{0.400pt}{11.322pt}}
\put(697.0,517.0){\rule[-0.200pt]{4.818pt}{0.400pt}}
\put(697.0,564.0){\rule[-0.200pt]{4.818pt}{0.400pt}}
\put(832.0,534.0){\rule[-0.200pt]{0.400pt}{11.081pt}}
\put(822.0,534.0){\rule[-0.200pt]{4.818pt}{0.400pt}}
\put(822.0,580.0){\rule[-0.200pt]{4.818pt}{0.400pt}}
\put(902.0,476.0){\rule[-0.200pt]{0.400pt}{16.381pt}}
\put(892.0,476.0){\rule[-0.200pt]{4.818pt}{0.400pt}}
\put(892.0,544.0){\rule[-0.200pt]{4.818pt}{0.400pt}}
\put(1306,677){\makebox(0,0)[r]{Bugg}}
\put(1350,677){\raisebox{-.8pt}{\makebox(0,0){$\Box$}}}
\put(856,530){\raisebox{-.8pt}{\makebox(0,0){$\Box$}}}
\put(1079,478){\raisebox{-.8pt}{\makebox(0,0){$\Box$}}}
\put(1243,467){\raisebox{-.8pt}{\makebox(0,0){$\Box$}}}
\put(1373,401){\raisebox{-.8pt}{\makebox(0,0){$\Box$}}}
\put(1328.0,677.0){\rule[-0.200pt]{15.899pt}{0.400pt}}
\put(1328.0,667.0){\rule[-0.200pt]{0.400pt}{4.818pt}}
\put(1394.0,667.0){\rule[-0.200pt]{0.400pt}{4.818pt}}
\put(856.0,473.0){\rule[-0.200pt]{0.400pt}{27.703pt}}
\put(846.0,473.0){\rule[-0.200pt]{4.818pt}{0.400pt}}
\put(846.0,588.0){\rule[-0.200pt]{4.818pt}{0.400pt}}
\put(1079.0,431.0){\rule[-0.200pt]{0.400pt}{22.645pt}}
\put(1069.0,431.0){\rule[-0.200pt]{4.818pt}{0.400pt}}
\put(1069.0,525.0){\rule[-0.200pt]{4.818pt}{0.400pt}}
\put(1243.0,420.0){\rule[-0.200pt]{0.400pt}{22.645pt}}
\put(1233.0,420.0){\rule[-0.200pt]{4.818pt}{0.400pt}}
\put(1233.0,514.0){\rule[-0.200pt]{4.818pt}{0.400pt}}
\put(1373.0,363.0){\rule[-0.200pt]{0.400pt}{18.067pt}}
\put(1363.0,363.0){\rule[-0.200pt]{4.818pt}{0.400pt}}
\put(1363.0,438.0){\rule[-0.200pt]{4.818pt}{0.400pt}}
\put(1306,632){\makebox(0,0)[r]{Siegel}}
\multiput(1328,632)(20.756,0.000){4}{\usebox{\plotpoint}}
\put(1394,632){\usebox{\plotpoint}}
\put(321,447){\usebox{\plotpoint}}
\multiput(321,447)(19.572,-6.908){6}{\usebox{\plotpoint}}
\multiput(423,411)(20.001,-5.545){5}{\usebox{\plotpoint}}
\multiput(524,383)(20.051,-5.360){5}{\usebox{\plotpoint}}
\multiput(625,356)(20.064,-5.311){5}{\usebox{\plotpoint}}
\multiput(727,329)(20.193,-4.798){5}{\usebox{\plotpoint}}
\multiput(828,305)(20.321,-4.225){5}{\usebox{\plotpoint}}
\multiput(929,284)(20.589,-2.624){5}{\usebox{\plotpoint}}
\put(1031,271){\usebox{\plotpoint}}
\put(321,615){\usebox{\plotpoint}}
\multiput(321,615)(18.194,-9.989){6}{\usebox{\plotpoint}}
\multiput(423,559)(19.165,-7.969){5}{\usebox{\plotpoint}}
\multiput(524,517)(19.551,-6.969){6}{\usebox{\plotpoint}}
\multiput(625,481)(19.387,-7.413){5}{\usebox{\plotpoint}}
\multiput(727,442)(19.786,-6.269){5}{\usebox{\plotpoint}}
\multiput(828,410)(20.051,-5.360){5}{\usebox{\plotpoint}}
\multiput(929,383)(20.204,-4.754){5}{\usebox{\plotpoint}}
\put(1031,359){\usebox{\plotpoint}}
\put(1306,587){\makebox(0,0)[r]{this work}}
\put(1328.0,587.0){\rule[-0.200pt]{15.899pt}{0.400pt}}
\put(321,473){\usebox{\plotpoint}}
\multiput(321.00,471.92)(2.148,-0.496){45}{\rule{1.800pt}{0.120pt}}
\multiput(321.00,472.17)(98.264,-24.000){2}{\rule{0.900pt}{0.400pt}}
\multiput(423.00,447.95)(44.891,-0.447){3}{\rule{27.033pt}{0.108pt}}
\multiput(423.00,448.17)(145.891,-3.000){2}{\rule{13.517pt}{0.400pt}}
\multiput(625.00,444.93)(11.795,-0.489){15}{\rule{9.122pt}{0.118pt}}
\multiput(625.00,445.17)(184.066,-9.000){2}{\rule{4.561pt}{0.400pt}}
\multiput(828.00,435.94)(29.579,-0.468){5}{\rule{20.400pt}{0.113pt}}
\multiput(828.00,436.17)(160.659,-4.000){2}{\rule{10.200pt}{0.400pt}}
\put(321,640){\usebox{\plotpoint}}
\multiput(321.00,638.92)(1.191,-0.498){83}{\rule{1.049pt}{0.120pt}}
\multiput(321.00,639.17)(99.823,-43.000){2}{\rule{0.524pt}{0.400pt}}
\multiput(423.00,595.92)(3.931,-0.497){49}{\rule{3.208pt}{0.120pt}}
\multiput(423.00,596.17)(195.342,-26.000){2}{\rule{1.604pt}{0.400pt}}
\multiput(625.00,569.92)(3.417,-0.497){57}{\rule{2.807pt}{0.120pt}}
\multiput(625.00,570.17)(197.175,-30.000){2}{\rule{1.403pt}{0.400pt}}
\multiput(828.00,539.92)(5.159,-0.496){37}{\rule{4.160pt}{0.119pt}}
\multiput(828.00,540.17)(194.366,-20.000){2}{\rule{2.080pt}{0.400pt}}
\end{picture}

figure 17
  \end{center}
\end{figure}
%
The solid lines band in fig.~17 shows the effect of adding our MEC
correction to the multiple scattering calculation of ref.~\cite{2}.
This correction is obtained by taking:
\begin{equation}
\Delta R^{\rm MEC}={\Delta\sigma^{\rm MEC}(^{12}C)\over 6 \sigma(^2
H)} \ ,
\label{A40}
\end{equation}
where  $\Delta\sigma^{\rm MEC}(^{12}C)=\sigma^{\rm IA+MEC}(^{12}C)-
\sigma^{\rm IA}(^{12}C)$, and $\sigma(^{2}H)$ has been obtained
directly from the Arndt phase-shifts by taking
$\sigma(^{2}H)=\sigma_n+\sigma_p$.
In calculating the MEC correction
we have taken only its contribution to the imaginary part of the
optical potential, while no correction has been done to the real part.
We observe that the addition of MEC correction to the conventional
optical calculation of Siegel {\it et al.} provides a significative
agreement of the calculation with the experiment for a large range of
energies.


\vspace{1cm}

\noindent
{\large \bf 7.- Conclusions}

\nobreak
\noindent
Let us summarize the results of our calculation. We have computed the
pionic cloud contribution to the $K^+$-nucleus selfenergy at lowest
order in $t_{K\pi}$ (see sections 4 and 5). Our result is well
described for $T_K<2.5~{\rm fm}^{-1}$ by:
\begin{equation}
\Pi^{\rm MEC}(T_K,{\bf r})=
B(T_K)\left ( {\rho({\bf r})\over \rho_0} \right )^2\ ,
\label{A41}
\end{equation}
where the coefficient $B(T_K)$ is
complex and $T_K$ is the kinetic energy of the incoming $K^+$.
We have found that, within our approximation, Re$B$ is energy
independent. Its precise value is not known due to the uncertainties
in the off-shell extrapolation of the $K\pi$ amplitudes (see Table 1).
On the other
hand, with very good approximation, Im$B(T_K)$ can be cast in the
following form:
\begin{equation}
{\rm Im}B(T_K)=\beta'_0~ \{ k_1~T_K~
+~k_2~(T_K-T_0)~\theta(T_K-T_0) \}\ ,
\label{A43}
\end{equation}
with $k_1=3.0~10^{-3}~{\rm fm}^{-2}$, $k_2=2.0~10^{-2}~{\rm fm}
^{-2}$, and $T_0=1.0~{\rm fm}^{-1}$. The constant $k_1$ is related to the
the diagram $d2$, and $k_2$ to the rest of the calculated MEC
diagrams $d_1,~d_3,~d_5$ which contribute to Im$B$ above a threshold
of around $T_0$ as
stated by the step function $\theta (T_K-T_0)$.
$\beta'_0$ is related to the $K^+\pi$ scattering lengths. For
the model of refs.~\cite{12,ref25}, one obtains
$\beta'_0=-8.1$~fm. The $K^+$-nucleus data for $^{12}C$ are
well reproduced for all energies by Re$B$=0,
$\beta'_0=-8.1$~fm, as can be seen in fig.~{17}, when conventional nuclear
corrections, taken from ref.~\cite{2}, are included together with
the MEC effects.

The main conclusion of this paper is that the effects of the mesonic
cloud in $K^+$ nucleus scattering are relevant and of the right order
of magnitude to account for the discrepancies of the conventional
optical potential \cite{2} with the
data. However, there are uncertainties, particularly in the real part
of the MEC, which do not allow us to draw stronger conclusions about
the actual size of the corrections. The main reason is the
sensitivity of the results to the off-shell extrapolation of the $KN$
scattering matrix for which there is not yet enough information.
Furthermore, we also noted that there are other sources of real part,
not linear in  the $KN$ $t-$matrix, which should also be considered.
Our evaluation of the imaginary part was however very precise within
the model of ref.~\cite{12} used here. However, there are also
approximations in Im$t_{K\pi}$ of ref.~\cite{12} since the parameter
$\beta'_0$ is tied to the threshold $K\pi$ amplitudes only.
Other models for the amplitude would also provide different results
for Im~$\Pi^{\rm MEC}$ although the restricted phase space for
Im~$\Pi^{\rm MEC}$ and the on-shell constraints in the amplitude make
this magnitude more stable.

The comparison of our results using Re~$\Pi^{\rm MEC}=0$ with the data
is very good.

On the other hand we also have observed that the direct relation of
$\Pi^{\rm MEC}$ to the distribution of excess pions in the nucleus,
as has been formerly assumed, is a consequence of a dangerous static
approximation which should be avoided. In the present case it
induced an error of a factor of two but in other cases it can induce
errors of several orders of magnitude. We also found that in any case
this contribution was only a small part of the total MEC corrections
tied to the interaction of kaons with the nuclear pions. These
findings should also serve as a warning for other calculations
directly relating the pion excess number to the modification of
nuclear magnitudes from the interaction of particles with the meson
cloud.

{\large \bf Acknowledgements: }

\nobreak
We thank to L.L. Salcedo, D.D. Strottman, M.J. Vicente-Vacas, W.
Weise and A. Steiner for helpful discussions and suggestions. Partial
financial support has been obtained from the following sources:
DGICYT (Spain) research project number PB92-0927 of the University of
Granada, CICYT (Spain) research project  AEN93-1205 of the University
of Valencia, Junta de Andalucia and INT University of Washington.

\vfill\eject

\noindent
{\large \bf Appendix A.-  The pion propagator}

\nobreak

We have taken
\begin{eqnarray}
D (q) = [{q^0}^2 - {\bf q}^{2} -
{m_\pi}^{2} - \Pi_\pi (q^{0},{\bf q}) ]^{-1} ,
 \label{Ee38}
\end{eqnarray}
\noindent
where
\begin{eqnarray}
\Pi_\pi (q) &=& {\bf q}^2{ \alpha(q)\over 1-g^\prime\alpha(q)}\ ,
\nonumber \\
\alpha(q)&=&\alpha_N(q)+\alpha_\Delta(q),\nonumber \\
\alpha_{N} (q) &=&
\left (\frac{f(q)}{m_\pi}\right )^{2} U_{N} (q),
\nonumber \\
\alpha_{\Delta} (q) &=& \left (
\frac{f (q)}{m_\pi}\right )^{2} U_{\Delta} (q), \nonumber\\
f(q)&=&f_\pi~F(q)~=~f_\pi~{\Lambda^2\over \Lambda^2+{\bf q}^2}\ ,
 \label{Ee39}
\end{eqnarray}
\noindent
where $1-g^\prime\alpha(q) $ is the Lorentz-Lorenz factor, with $g'$
the Landau-Migdal parameter, and $U_{N}, U_{\Delta}$, the Lindhard
functions for $ph$ and $\Delta h$ excitation with the appropriate
normalization are given explicitly in the appendix of \cite{ref29}.
$U_{\Delta} (q)$ has an imaginary part coming from $\Delta$ decay
into $\pi N$. We have taken the form factor static because the
relevant momenta involved in the process are well below $\Lambda$ and
the form factor does not play much of a role. This was already
investigated in \cite{12}. Not having  $q^{0}$ dependence in $F (q)$
simplifies the analytical structure and allows us to make the formal
developments of the former sections. We take $g' = 0.6$, $f_\pi^{2}/
4\pi = 0.08$, and  $\Lambda= 1250$ MeV.

The leading order in density of $\Pi^{(1)}_\pi$ is given by:
\begin{eqnarray}
\Pi^{(1)}_\pi=\Pi_{N}+\Pi_{\Delta}\ , \
\quad\quad \Pi_{N}(q)={\bf q}^2\alpha_N^{(1)}\ ,
\quad\quad \Pi_{\Delta}(q)={\bf q}^2\alpha_\Delta^{(1)}\ ,
\nonumber
\end{eqnarray}
where $\alpha_N^{(1)}$ and $\alpha_\Delta^{(1)}$ are the linear part
in density of $\alpha_N$ and $\alpha_\Delta$ of eq.~(\ref{Ee39})
Hence, the leading orders in density expansion for the pion
propagator $D(q)$ are:
\begin{eqnarray}
D_{(1)}(q)& = &
D_0^2(q)\Pi_\pi^{(1)}=D_{(1)}^{ph}+D_{(1)}^{\Delta h}\
,\nonumber \\
D_{(1)}^{ph}(q)& = & D_0^2(q)\Pi_N(q)\ ,
\quad\quad D_{(1)}^{\Delta h}(q)= D_0^2(q)\Pi_\Delta(q)\ ,\nonumber\\
D_{(2)}(q)& = &D_0^2(q){\bf q}^2\alpha^2(q)
(g^\prime+{\bf q}^2 D_0(q))\ ,
\nonumber
\end{eqnarray}

\vspace{1cm}

\noindent
{\large \bf Appendix B.- The $K^{+} \pi$ amplitude.}

\nobreak

As discussed before we need only the isoscalar $ K \pi$ amplitude
which we take from \cite{12}. This amplitude incorporates analytical
properties, unitarity, crossing symmetry $t^{0} (k',q';k,q) = t^{0}
(k', -q; k, -q')$,  and on-shell constrains and keeps up to linear
terms in $s$ and $u$ for forward scattering. Its justification and
uncertainties are clearly discussed in \cite{12} and we refer the
reader to this paper. The model contains an $s$-wave part and a
$p$-wave part. The $p$-wave part is accounted for by means of a I =
1/2 resonance, $K^{*} (892)$. After the implementation of the
crossing symmetry the two pieces give rise to the following isoscalar
amplitude.
\begin{eqnarray}
t^{0}  (k,q;k,q) &=& t^{(0,s)} (k,q;k,q) + t^{(0,p)} (k,q;k,q) ,
 \label{Ee40}\\
t^{(0,s)} (k,q;k,q) &=& \alpha_{0} + \beta_{0} (s + u) + i \beta'_{0}
[k (s) + k (u)] ,
 \label{Ee41}\\
t^{(0, p)} (k,q;k,q) &=&
T_{\rm res} (s) \theta (s - x_{0}) + T_{\rm res} (u) \theta
(u -x_{0}) ,
 \label{Ee42}
\end{eqnarray}
\noindent
with
\begin{eqnarray}
T_{\rm res}(x) &=&
- \frac{8 \pi \sqrt{x}}{K (x)} \frac{ M_{r} \Gamma (
K (x))}{M_{r}^{2} - x - i M_{r} \Gamma (K (x))} \ ,
 \label{Ee43}\\
\Gamma (K) &=& (\frac{K}{K_{r}})^{3}
\Gamma_{r} [ \frac{1 + (K_{r} R)^{2}}
{1 + (K R)^{2}} ] \ ,
 \label{Ee44}
\end{eqnarray}
\noindent
$x_{0} = (m_{K} +
{m_\pi})^{2}$ and $k (x), K (x)$ the CM momentum given by
\begin{eqnarray}
k (x) &=& \frac{\sqrt{
{m_\pi} m_{K}}}{
{m_\pi} + m_{K}} \sqrt{x - (
{m_\pi} + m_{k})^{2}} \ ,
 \label{Ee45}\\
K (x) &=& \left[ \frac{[x - (
{m_\pi} + m_{K})^{2}][x - (
{m_\pi} - m_{K})^{2}]}
{4x} \right]^{1/2}
 \label{Ee46}
\end{eqnarray}

\noindent
and the parameters: $\beta'_{0}=- 8.1$~fm, $M_{r}$=895.7~MeV,
$\Gamma_{r}$=52.9~MeV, \hfill\break $R$=4.3~(GeV/$c$)$^{-1}$, $K_{r}$
given by eq.~(\ref{Ee46}) for $x = M_{r}$. Three different
parametrizations I, II and III, with values for  $\alpha_{0},~
\beta_{0}$, given in table~1 are taken from \cite{12}. These
parameters are constrained by the relationship $
\alpha_0+2(m_\pi^2+m_K^2)\beta_0=-11.0$, which follows from
eq.~(\ref{Ee41}) on-shell and at threshold. Notice that
parametrization II does not fulfill this requirement. So we will not
use it.

Note that $K (x)$ is the relativistic CM momentum while $k (x)$
involves a nonrelativistic approximation. The choice of  $k (x)$ for
the $s$-wave part is done in \cite{12} to avoid extra singularities
and is consistent with some threshold approximations involved in the
derivation of eq.~(\ref{Ee41}).

Since $T_{\rm res} (x)$ is zero below pion threshold there are no
problems in separating the real and imaginary parts. However, the
imaginary part in the $s$-wave term gives rise to a real part below
pion threshold when extrapolated analytically. Thus
\begin{eqnarray}
 \, t^{(0,s)} (k,q;k,q) =  \, \tilde{t}\,^{(o,s)} (s) +
 \, \tilde{t}\,^{(o,s)}
(u)
 \label{Ee47}
\end{eqnarray}
\noindent
with
\begin{eqnarray}
{\rm Re} \, \tilde{t}\,^{(o,s)} (x)& =& \left\{
\begin{array}{ll}
\frac{\alpha_{0}}{2} + \beta_{0} x & , x > x_{0} \\
\frac{\alpha_{0}}{2} +
\beta_{0} x - \beta'_{0} k_{I} (x)  &, x < x_{0}
\end{array}\right.\nonumber\\
{\rm Im} \, \tilde{t}\,^{(o,s)} (x)& =& \left\{
\begin{array}{ll}
 \beta'_{0} k(x)  &, x > x_{0} \\
0  &, x <  x_{0}
\end{array}\right.
 \label{Ee48}
\end{eqnarray}
\noindent
with
\begin{eqnarray}
k_{I} (x) = \frac{\sqrt{
{m_\pi} m_{K}}}{
{m_\pi} + m_{K}} \sqrt{x_{0} - x} .
 \label{Ee49}
\end{eqnarray}
Coming back to eqs.~(\ref{Ee28}) and~(\ref{Ee29}),
we discussed that in the general case for $u>(m_\pi+m_K)^2$.
\begin{eqnarray}
\lefteqn{\sum_{l} \sum_{i,j} | t_{K^{+} \pi^{l} \rightarrow  K^{i}
\pi^{j}} ( k',q';k, -q)
|_{\rm ave}^{2} = - \frac{ 8 \pi \sqrt{u}}{K (u)} 3
{\rm Im}~\tilde{t} (u)}
\label{Ee50} \\
 & & = - 24\pi \left\{  \sqrt{u} \frac{k (u)}{K (u)} \beta'_{0} -
{\rm Im}~
\frac{8 \pi u}{K^{2} (u)} \frac{ M_{r} \Gamma (K (u))
\theta (u - x_{0})}{M_{r}^{2} - u - iM_{r} \Gamma (K (u))} \right\}
 \label{Ee51} ,
\end{eqnarray}
with $u = (k - q)^{2}$.
However, the threshold approximations involved in the s-wave part in
\cite{12} require that all factors in the s-wave part in
eq.~(\ref{Ee51}) be calculated at pion threshold. We have checked
that this induces about 5$\%$ differences in the evaluation of the
diagram of fig.~{2a} with respect to the results  keeping the $u$
dependent factors. Observe that eq.~(\ref{Ee51}) provides the
analytical continuation of the the left hand side of eq.~(\ref{Ee50})
for $u<(m_\pi+m_K)^2$.

Note that if the factor $\sqrt{u}$ is kept we run into problems if
the line $q'$ excites a $ph$, like in diagrams $d1$, $d2$, $d3$ of
fig.~{9}, since $u$ can become negative and this leads to the absurd
conclusion that $|t^{2}|$ is purely imaginary. The threshold
approximations done in \cite{12} aimed at avoiding such pathologies.

Thus, taking this into account this average we can rewrite
eq.~(\ref{Ee23}) in the form:
\begin{eqnarray}
{\rm Im} \tilde\Pi (k) & = &
2 \int \frac{d^{4} q}{(2 \pi)^{4}} \int \frac{d^{4} q'}
{(2 \pi)^{4}}\theta (q^{0}) \theta ( q'^{0})
\theta (k^{0} - q^{0} - q'^{0})
\nonumber \\
& \times &
 {f} (u)  {\rm Im} D (q)
{\rm Im} D (q') {\rm Im} D_{K} (k - q - q') ,
 \label{Ee52}
\end{eqnarray}
\begin{eqnarray}
{f} (u) = - 24\pi \left\{ \beta'_{0}
(m_\pi+m_K) - \frac{8 \pi u}{K^{2}
(u)}
{\rm Im}\frac{ M_{r}
\Gamma (K (u)) \theta (u- x_{0})}{M_{r}^{2} - u - i
M_{r} \Gamma (K (u))} \right\} .
 \label{Ee53}
\end{eqnarray}
For the $s-$wave this is equivalent to consider the on-shell values
for the $t_{K\pi}-$matrix when calculating $|t|^2$ but evaluated at
threshold.
\vspace{1cm}

\noindent
{\large \bf Appendix C.- The impulse approximation}

\nobreak

Our $K^{+}$ selfenergy in the impulse approximation is given by
\begin{eqnarray}
\Pi^{\rm (IA)} (k) = -
\frac{4 \pi \sqrt{s}}{M} \left\{ \frac{1}{2} (f_{0} + f_{1})
\rho_{n} ({\bf r}) + f_{1} \rho_{p} ({\bf r}) \right\} ,
 \label{Ee54}
\end{eqnarray}
\noindent
with M the nucleon mass, where $f_{I}$ are the $K^{+} N$ spin non
flip isospin amplitudes and $\rho_{n} ({\bf r}), \rho_{p} ({\bf r})$
the neutron and proton densities.

We consider $s$, $p$ and $d$ waves. For the $p$-wave we substitute
\begin{eqnarray}
{\bf q} . {\bf q}\,' \rho ({\bf r}) \rightarrow
- \frac{M^{2}}{s} {\bf \nabla}
\rho ({\bf r}\,) \, {\bf \nabla} ,
 \label{Ee55}
\end{eqnarray}
\noindent
as customarily done in pionic atoms \cite{ref30}, and for the
$d$-wave, which is almost negligible in the range that we study, we
take only the forward value in eq.~(\ref{Ee54}).

The $K^{+} N$ phase shifts are taken from \cite{ref27}. The kaon
selfenergy in the impulse approximation is shown in figs.~{5} and
{13} for nuclear matter density.
\newpage
{\bf Table 1}

\vspace{1cm}

\begin{tabular}{|r||r|r|r|c|c|}
\hline
 & $\alpha_0$ & $\beta_0\ \ $ &$\beta'_0\ \ $
 & {\rm Re}$\delta\Pi_{\rm stat}$($\rho=\rho_0)$
 & {\rm Re}$\delta\Pi$($\rho=\rho_0$) \\
 & & & & &  $\equiv{\rm Re}B$ \\
 & &  [fm$^2$] & [fm] & [fm$^{-2}$]\ \ \ \ \ \
 & [fm$^{-2}$]\ \ \ \ \
\\ \hline \hline
I   & \  18.7 & $-$2.2 & $-$8.1 & \  0.54 & \  0.58 \\ \hline
II  & \  11.4 & $-$1.0 & $-$8.1 & \       & \       \\ \hline
III & $-$11.0 & \  0.0 & $-$8.1 & $-$0.22 & $-$0.22 \\ \hline
IV  &$-$\ 2.8 & $-$0.6 & $-$8.1 & $\sim$0.00 & \  0.00 \\ \hline
\end{tabular}

\begin{itemize}

\item {\bf Table 1:}
Different parametrizations for the $K-\pi$
$t$-matrix, and the results of the real part of the $K^+$
optical potential due to the pionic cloud for normal nuclear
matter for each parametrization.

\end{itemize}

\newpage
\centerline{Figures captions}

\begin{itemize}
\item {\bf Figure  1:} ${\delta N({\bf q})}$ versus $q$ is shown for
normal nuclear matter.

\item {\bf Figure  2:} ``Standard'' $K^+$ selfenergy diagrams due to
the pion cloud.

\item {\bf Figure  3:}  $K^+$ scattering with a pion, annihilating
two pions from the ground state or creating two pions from the ground
state.

\item {\bf Figure  4:} The complex plane $q^0$ with the cuts and
poles of $D(q)\tilde t(u)$. The adequate Wick rotation for
performing its integration in $q^0$ is shown.

\item {\bf Figure  5:} Imaginary part of the $K^+$ selfenergy for
normal nuclear matter versus kinetic energy of the incoming kaon.

\item {\bf Figure  6:} The optical theorem for the $K\pi$ amplitude
is diagrammatically shown.

\item {\bf Figure  7:} Feynman diagram which imaginary part is
related, through the optical theorem, to the imaginary part of the
``standard'' diagram of fig.~{2}.

\item {\bf Figure  8:} This diagram contains all the contributions
to the imaginary part of the $K^+$ selfenergy up to second order in
$t_{K\pi}$.

\item {\bf Figure  9:} Diagrams $d1$, $d2$, $d3$ and  $d4$. They
contribute to Im$\Pi^{\rm MEC}$ up to second order in density.

\item {\bf Figure  10:}  Imaginary part of the $K^+$ selfenergy
versus density for 450 MeV of kinetic energy of the incoming kaon for
diagrams  $d1$ and  $d3$ are shown.  The lines are exact quadratic
functions in density for comparison.

\item {\bf Figure 11:} Diagram $d5$. It contributes to Im$\Pi^{\rm
MEC}$ in second order in density.

\item {\bf Figure 12:}  Imaginary part of the $K^+$ selfenergy for
normal nuclear matter versus kinetic energy of the incoming kaon from
the different mechanisms considered.

\item {\bf Figure 13:}  Real part of the $K^+$ selfenergy for normal
nuclear matter versus kinetic energy of the incoming kaon.

\item {\bf Figure 14:} Differential cross section of a $K^+$ with
kinetic energy of 450~MeV scattered by a $^{12}$C nucleus. Dashed
line depicts the results using the IA. Long-dashed (dot-dashed) line
includes IA plus the the optical potential coming from the pionic
cloud using parametrization I (III). The experimental data are from
ref.~\cite{5}.

\item {\bf Figure 15:} Same as fig.~{14} for $^{40}$Ca. The
experimental data are from ref.~\cite{6}.

\item {\bf Figure 16:} a) Total cross section for $K^+$ scattered by
$^{12}$C versus kinetic energy of the $K^+$.
b) The ratio $R={\sigma(K^+~^{12}C)\over 6\sigma(K^+~^{2}H)}$
versus the lab momentum of the $K^+$.
Dashed line corresponds to IA. Solid line (IV) includes IA
plus the imaginary part of the optical potential coming from the
pionic cloud. Dotted lines includes IA plus the imaginary and real
parts of the optical potential coming from the pionic cloud for two
different parametrizations (I, III) of the $t_{K\pi}$ amplitude. The
experimental data are: crosses from ref.~\cite{8}, diamonds from
ref.~\cite{4}, squares from
ref.~\cite{6}.

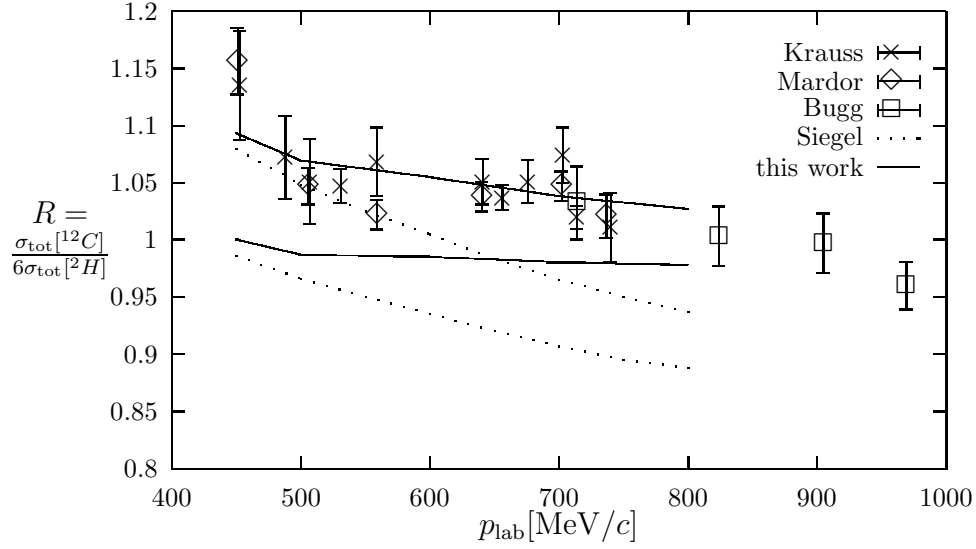
\item {\bf Figure 17:} The ratio of cross-sections
$R={\sigma(K^+~^{12}C)\over 6\sigma(K^+~^{2}H)}$ is plotted against the
lab momentum of the incoming $K^+$. The experimental data are:
crosses from ref.~\cite{8}, diamonds from ref.~\cite{4}, squares from
ref.~\cite{6}. The two dashed lines define the band of uncertainties
of the theoretical results for $R$ obtained with the conventional
optical potential of ref.~\cite{2}.  The two solid lines define
the band of uncertainties of the theoretical results for $R$ when the
corrections due to MEC calculated here are added to the results
obtained with the conventional optical potential of ref.~\cite{2}.

\end{itemize}

paper has 17 FIGURES, only fig.17 is included in this LaTeX file \\
please, REQUEST THE FIGURES to C. Garcia-Recio, e-mail: G\_RECIO@UGR.ES
\\


\newpage
\begin{thebibliography}{99}
\bibitem{1} P. B. Siegel, W. B. Kaufmann and W. R. Gibbs,
Phys. Rev. C {\bf 30}, 1256 (1984).
\bibitem{2} P. B. Siegel, W. B. Kaufmann and W. R. Gibbs,
Phys. Rev. C {\bf 31}, 2184 (1985).
\bibitem{3} C. M. Chen and D. J. Ernst, Phys. Rev. C {\bf 45},
2019 (1992).
\bibitem{4} D. V. Bugg et al., Phys. Rev.{\bf 168}, 1466 (1968).
\bibitem{5} D. Marlow et al., Phys. Rev. C {\bf 25}, 2619 (1982).
\bibitem{6} Y. Mardor et al., Phys. Rev. Lett. {\bf 65}, 2110 (1990).
\bibitem{7} J. Alster, Proc. if the Int. Conference on
Hypernuclear and Strange
Particle physics, Shimoda, Japan, 1991 [Nucl. Phys. {\bf A547}, 321c
(1992)].
\bibitem{8} R. A. Krauss et al., Phys. Rev. C {\bf 46}, 2019 (1992).
\bibitem{9} G. E. Brown, C. B. Dover, P. B. Siegel and W. Weise,
Phys. Rev. Lett. {\bf 60}, 2723 (1988).
\bibitem{10} W. Weise, Nuovo Cimento {\bf 102A}, 265 (1989).
\bibitem{11} S. V. Akulinichev, Phys. Rev. Lett.
{\bf 68}, 290 (1992).
\bibitem{12} M. F. Jiang and D. S. Koltun,
Phys. Rev. C {\bf 46}, 2462 (1992).
\bibitem{13} B. R. Martin, D. Morgan and G. Shaw,
Pion- Pion Interaction in Particle Physics
(Academic Press, New York, 1976) chap 14.1 and 14.3.
\bibitem{14} E. L. Berger, F. Coester and R. B. Wiringa, Phys. Rev.
D {\bf 29}, 398 (1984); E. L. Berger and F. Coester, Phys. Rev.
D {\bf 32}, 1071 (1985).
\bibitem{ref15} B. L. Friman, V. R. Pandharipande and R. B.
Wiringa, Phys. Rev. Lett {\bf 51},  763 (1983).
\bibitem{ref16} J. Nieves, E. Oset and C. Garc\'{\i}a-Recio,
Nucl. Phys. {\bf A554}, 509 (1993); $ibid$ pag. 554.
\bibitem{ref17} C. Itzykson and J. B. Zuber, Quantum Field Theory,
McGraw Hill, 1985.
\bibitem{ref18} M. Ericson and M. Rosa-Clot,
Phys. Lett. B {\bf 188}, 11 (1987).
\bibitem{ref19} A. D. Martin and T. D. Spearman,
''Elementary Particle Theory'', North-Holland, 1970, Amsterdam.
\bibitem{ref20} P. Fern\'andez de C\'ordoba and E. Oset, Nucl. Phys.
{\bf A554}, 793 (1992).
\bibitem{ref21} H. Bando and H. Takaki, Phys. Lett. B {\bf 150}
, 409 (1985).
\bibitem{ref22} O. Meirav, E. Friedman, R. R. Johnson, R.
Olszewski and P. Weber, Phys. Rev. C {\bf 40}, 843 (1989).
\bibitem{ref23} C. Garc\'{\i}a-Recio, J.M. Nieves and E. Oset,
Act. Phys. Polonica B {\bf 24}, 1757 (1993).
\bibitem{ref24} E. Friedman,
Act. Phys. Polonica B {\bf 24}, 1673 (1993).
\bibitem{ref25} P. Estabrooks {\it et al.}, Nucl. Phys. B133,  490
(1978).
\bibitem{ref26} B. Martin, Nucl. Phys. {\bf B94}, 413 (1975).
\bibitem{ref27} R. A. Arndt, L. D. Roper and P. II. Steinberg,
Phys. Rev.D {\bf 18}, 3278 (1978);
R. A. Arndt and L. D. Roper, Phys. Rev. D {\bf 31}, 2230 (1985).
\bibitem{ref28} Q. Daffner, Diplomarbeit, University of
Regensburg, 1990; W. Weise, Private communication.
\bibitem{ref29} E. Oset, P. Fern\'andez de C\'ordoba,
L. L. Salcedo and R. Brockmann, Phys. Reports {\bf 188}, 79 (1990).
\bibitem{ref30} M. Ericson and T.E.O. Ericson,
Ann. Phys. (N.Y.) {\bf 36}, 323 (1966).
\end{thebibliography}
\end{document}